\documentclass[aps,prl,reprint,superscriptaddress,floatfix,nobibnotes]{revtex4-2}
\usepackage[T1]{fontenc}
\usepackage{amsmath,amssymb,mathtools,amsthm}
\mathtoolsset{showonlyrefs}
\usepackage{graphicx}
\usepackage{booktabs}
\usepackage{times}
\usepackage{xcolor}
\usepackage[hypertexnames=false,colorlinks=true,citecolor=teal,linkcolor=teal,urlcolor=teal]{hyperref}
\usepackage{tikz-cd}
\usetikzlibrary{backgrounds}

\newcommand{\code}[1]{[\![#1]\!]}

\theoremstyle{plain}
\newtheorem{lemma}{Lemma}

\begin{document}

\title{Parallel algebraic surgery for qLDPC codes with constant shuttling depth}

\author{Tian-Gang Zhou}
\thanks{Source code and interactive shuttling animations accompanying this work are available in the
  \href{https://github.com/tiangangzhou/hardware-efficient-surgery}{GitHub repository}
  and on the
\href{https://tiangangzhou.github.io/hardware-efficient-surgery/lp-surgery-aod-polynomials.html}{project webpage}, resp.}
\affiliation{Department of Quantum Matter Physics, University of Geneva, 1205 Geneva, Switzerland}
\author{Boren Gu}
\affiliation{Dahlem Center for Complex Quantum Systems, Freie Universit\"at Berlin, 14195 Berlin, Germany}
\author{Jens Eisert}
\email{jense@zedat.fu-berlin.de}
\affiliation{Dahlem Center for Complex Quantum Systems, Freie Universit\"at Berlin, 14195 Berlin, Germany}
\author{Chen Zhao}
\email{chenzhao@eitech.edu.cn}
\affiliation{School of Mathematical Sciences, Eastern Institute of Technology, Ningbo, 315200 Ningbo,  China}

\begin{abstract}
  Reconfigurable atom arrays provide a promising platform for high-rate quantum low-density parity-check codes, but their computational advantage depends on whether logical operations can be implemented without incurring prohibitive shuttling and space overheads. We introduce algebraic surgery for lifted product codes, co-designing the logical measurements, auxiliary codes, and atom shuttling to enable hardware-efficient Pauli-product measurements. These protocols provide flexible logical addressability subject to the cyclic-symmetry constraint, enabling a broad class of parallel logical measurements, including canonical logical operators with heavily overlapping physical supports and joint measurements within or between code blocks even with different lift sizes, while achieving constant shuttling depth on neutral-atom processors under explicit layout assumptions. In a representative hardware-model benchmark, a cycle supporting 33 parallel canonical logical measurements is about $2\times$ faster than the compared single-observable graph-based surgery schedules, with competitive qubit overhead. These shuttling and circuit-level simulations confirm that the algebraic structure of quantum low-density parity-check codes can be used not only to reduce space overhead, but also to simplify physical implementation and expand the set of efficiently accessible logical operations on neutral-atom processors.
\end{abstract}
\maketitle

\paragraph{Introduction.}
\label{sec:intro}
Neutral-atom quantum processors combine global laser control, which enables highly parallel entangling gates, with reconfigurable connectivity achieved by atom shuttling~\cite{LukinQEC,LukinFaultTolerant,PlanQCMeasurementFree,RodriguezRobinsonJepsen2025,5d8p-3hm1}.
These capabilities open possibilities for
high-rate encodings~\cite{Xu2024ConstantOverhead,zhao2026ultrahighratequantumerrorcorrection,Preskill10000} and efficient logical gates~\cite{LukinQEC,LukinFaultTolerant,cain2024correlateddecoding,zhou2025lowoverheadtransversal}, but shuttling atoms through hardware-constrained geometries is not free: the shuttling trajectories and scheduling contribute directly to the cost of fault-tolerant computation. This tension is especially important in high-rate quantum architectures, where recent resource estimates~\cite{Preskill10000,Pinnacle,TripierChungYoung2026,MindTheGaps} emphasize low qubit overhead without directly accounting for shuttling efficiency, leaving the practical cost of logical operations an urgent question: \textit{Can the parallelism and reconfigurable connectivity of neutral-atom processors be converted into logical operations with practical shuttling requirements?}

Recent developments in \emph{quantum low-density parity-check} (qLDPC) codes provide a foundation for high-rate fault-tolerant architectures. Beyond their asymptotic guarantees of non-vanishing encoding rates and growing distances~\cite{PRXQuantum.2.040101,9996782,Panteleev2021degeneratequantum,bravyi2024high,ScrubyHillmannRoffe2026}, good finite-size instances with ultra-high encoding rates or distances have been proposed and evaluated~\cite{zhao2026ultrahighratequantumerrorcorrection,Preskill10000,yang2026gala,BhardwajMaMeister2026,lu2026ultralowoverhead,LeeOkadaMaskara2026,LiangGuChen2026}. The algebraic structure of these codes has also motivated efficient hardware implementations of quantum memories based on finite-size code instances~\cite{BhardwajMaMeister2026,zhao2026ultrahighratequantumerrorcorrection,yang2026gala,OurPlanarArchitecture,yang2025planarfaulttolerantquantumcomputation}. However, hardware efficiency in the memory does not automatically extend to logical operations.

For qLDPC architectures, code surgery offers one of the main frameworks for realizing such logical operations through logical \textit{Pauli-product measurements} (PPMs) that can combine low space-time overhead with flexible addressability and parallelism~\cite{HorsmanSurgery,IdeGowda2025,ZhengJiangXu2025HighRateSurgery,CowtanHeWilliamson2026,he_extractors_2025,CohenKimBartlett2022,WilliamsonYoder2026,ZhengZhengJiangXu2026CanonicalLP,BhardwajMaMeister2026,QGPU,DiFiniIcebergSurgery,BlueHeZhou2026,LogicalMeasurementsQLDPC}. In a typical construction, an auxiliary code or chain complex is attached to the data code to form a merged code, and products of auxiliary checks in an appropriate kernel induce the desired logical PPM on the data block~\cite{IdeGowda2025}. Existing high-rate and parallel constructions often realize this auxiliary connectivity through graph-based or other unstructured designs~\cite{IdeGowda2025,WilliamsonYoder2026,he_extractors_2025,ZhengJiangXu2025HighRateSurgery}. Such constructions offer considerable flexibility in selecting logical observables, but the auxiliary connectivity need not preserve the algebraic structure of the underlying data code. This flexibility can therefore come at a hardware cost: when the regular structure of the data code is lost in surgery, the resulting logical operations need not inherit its shuttling efficiency. A general construction that preserves the hardware advantages of structured qLDPC codes remains an open challenge.

\begin{figure*}[t]
  \centering
  \includegraphics[width=\textwidth]{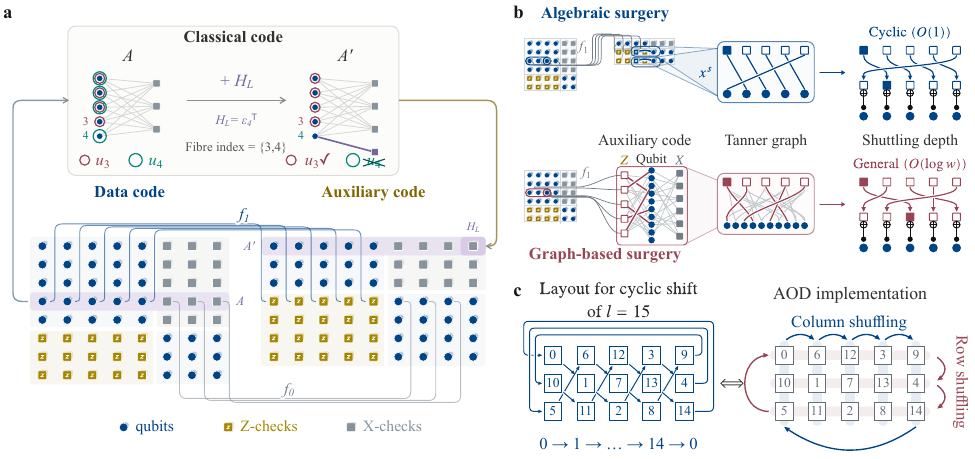}
  \caption{\textbf{Algebraic surgery and its hardware implementation.}
    (a) A cyclic LP code can be defined on a 2D grid of data/check patches, each with $l$ circulant physical qubits. The surgery connections $f_0$ and $f_1$ on the data and check qubits are shown between the data and auxiliary codes. We highlight two rows in LP blocks to illustrate the core mechanism of our algebraic surgery. The original $A$ supports two logical fibre generators, $u_3$ and $u_4$, indicated by circles of two different colors, while the additional $H_L$ eliminates $u_4$, which has support on index $4$.
    (b) A monomial edge $x^s$ in the auxiliary protograph lifts to a cyclic Tanner-graph matching, whereas the illustrated graph-based auxiliary construction gives a general matching. Each matching is implemented by permuting ancilla qubits and applying parallel entangling gates, illustrated here as CNOTs. Let $w$ denote the weight of the measured logical operator. Under the shuttling assumptions described in the text, the corresponding shuttling depths are $O(1)$ and $O(\log w)$, respectively.
  (c) For $l=15$, the integer-CRT layout $a\mapsto(a\bmod 3,a\bmod 5)$ realizes a cyclic shift through efficient AOD column and row shuffling.}
  \label{fig:overview}
\end{figure*}

We address these challenges by introducing \textit{algebraic surgery} for \textit{lifted product} (LP) codes defined over $\mathbb F_2$ group algebras of odd-order cyclic groups~\cite{9567703,10.1109/TIT.2021.3097347}. The surgery
is constructed directly at the group-algebra level, before expansion into binary matrices, using an auxiliary LP code attached to the
data code through an injective coordinate map. This construction preserves the cyclic structure that supports efficient atom shuttling. The construction also enables parallel logical measurements subject to a cyclic-symmetry constraint, providing access to complete cyclic \textit{logical fibres}~\cite{ZhengZhengJiangXu2026CanonicalLP,Lee2026LogicalSpectroscopy} or cyclically invariant subspaces of \textit{logical packets} within each fibre. Parallel PPMs can be performed within a block or between blocks, even with different lift sizes. We derive explicit conditions for logarithmic and constant shuttling depth in terms of the underlying protograph, and benchmark the resulting elapsed times against recent graph-based surgery constructions~\cite{WilliamsonYoder2026,IdeGowda2025} under hardware specifications informed by recent neutral-atom experiments~\cite{LukinQEC,Xu2024ConstantOverhead}. Explicit shuttling-time simulations show that algebraic surgery has a cycle time close to that of the memory, whereas graph-based schedules require roughly twice the memory cycle time. Circuit-level simulations give memory-comparable \emph{logical error rates}  (LERs).

The remainder of the paper is organized as follows. The section on algebraic surgery and hardware implementation presents the construction of algebraic surgery and compares the shuttling depth of algebraic and graph-based surgery. The section on addressability of algebraic surgery develops an addressing framework based on logical fibres in canonical bases, \emph{Chinese-remainder-theorem} (CRT) packets, and adapters for connecting fibres. The section on numerical simulation presents the shuttling-time and LER simulations. Finally, we discuss potential applications and future directions.

\paragraph{Algebraic surgery and hardware implementation.}
\label{sec:construction-and-complexity}

We define algebraic surgery as a structured specialization of \emph{Calderbank-Shor-Steane} (CSS) code surgery in which the data and auxiliary codes are both LP codes and their attachment is specified over the underlying group algebra by a coordinate map that is injective on the selected auxiliary kernel. This keeps the surgery construction within the same algebraic family as the data code, rather than introducing an arbitrary auxiliary connectivity graph.

Consider a CSS data code~\cite{CalderbankShor,Steane1996} $C$ of distance $d$ with checks $H_X,H_Z$, and an auxiliary CSS complex with boundary maps $\partial_1,\partial_0$. Write $\dagger$ for binary transpose. A compatible attachment $f=(f_1,f_0)$ satisfies $H_Xf_1=f_0\partial_1$, with $\partial_0\partial_1=0$, and gives the merged checks~\cite{IdeGowda2025}
\begin{equation}
  \widetilde H_Z=
  \begin{pmatrix}
    H_Z&0\\
    f_1^\dagger&\partial_1^\dagger
  \end{pmatrix},
  \qquad
  \widetilde H_X=
  \begin{pmatrix}
    H_X&f_0\\
    0&\partial_0
  \end{pmatrix}.
  \label{eq:css-surgery-checks}
\end{equation}
Products of the added $Z$ checks indexed by $\xi\in\ker\partial_1$ cancel on the auxiliary system, leaving the data-code action $f_1\xi$. The data-code logical $Z$ classes form $H_1(C)=\ker H_X/\operatorname{im}H_Z^\dagger$. Hence the logical subspace measured by the surgery is
\begin{equation}
  M=[f_1\ker\partial_1]\subseteq H_1(C),
  \label{eq:css-surgery-readout}
\end{equation}
where $[\cdot]$ means logical class. Preparing the auxiliary qubits in $\lvert+\rangle$, extracting the merged checks, and finally measuring the ancilla $X$ qubits implement the corresponding logical $Z$ measurements, with classical post-processing and Pauli-frame updates. To retain the protection of the original data code, we require the merged code to remain LDPC with distance
at least $d$ and repeat merged-check extraction for $d$ rounds~\cite{IdeGowda2025}. $X$ logical measurements can
be carried out simply by exchanging $X$ and $Z$.

For an odd lift size $l$, set $R_l=\mathbb F_2[x]/(x^l+1)$~\cite{9567703,QGPU}. Over $R_l$, $\dagger$ combines transpose with $x\mapsto x^{-1}$. For algebraic surgery, we take the data code to be $C=\mathrm{LP}(A,B)$ over $R_l$. Factor maps $A:R_l^{n_A}\to R_l^{m_A}$ and $B:R_l^{n_B}\to R_l^{m_B}$ define an LP code with $n=l(n_A m_B+m_A n_B)$ physical qubits. With tensor products over $R_l$, its check matrices are
\begin{equation}
  \begin{aligned}
    H_X&=(A\otimes I_{m_B}\mid I_{m_A}\otimes B),\\
    H_Z&=(I_{n_A}\otimes B^\dagger\mid A^\dagger\otimes I_{n_B}).
  \end{aligned}
  \label{eq:main-lp-checks}
\end{equation}
Let $A':R_l^{n_A}\to R_l^{m_A+s}$ augment $A$ with $s$ logical-selection rows, and let $D:R_l^{n_D}\to R_l^{m_D}$ be a second factor. We choose the auxiliary code itself to be another LP code $\mathcal A=\mathrm{LP}(A',D)$. This construction follows the thickening used in product surgery~\cite{CohenKimBartlett2022,ZhengZhengJiangXu2026CanonicalLP}. Its boundary maps are
\begin{equation}
  \partial_1=
  \begin{pmatrix}
    A'\otimes I_{n_D}\\
    I_{n_A}\otimes D
  \end{pmatrix},
  \qquad
  \partial_0=
  \bigl(I_{m_A+s}\otimes D\mid A'\otimes I_{m_D}\bigr).
  \label{eq:main-tube}
\end{equation}
Choose $D$ with $\ker D=R_lt$ and normalize a unit coordinate to $t_q=1$. Then $\ker\partial_1=\ker A'\otimes R_lt$, preserving the selected kernel. We assess each merged-code distance separately.

For a selected output coordinate $j$ of $B$, let $\varepsilon_j\in R_l^{m_B}$ and $\varepsilon_q\in R_l^{n_D}$ be standard coordinate vectors, and set $W=\varepsilon_j\varepsilon_q^{\mathsf T}$ and $P_A=(I_{m_A}\mid0)$. The coordinate attachment is
\begin{equation}
  f_1=\binom{I_{n_A}\otimes W}{0},\qquad
  f_0=(P_A\otimes W\mid0).
  \label{eq:main-attachment}
\end{equation}
Since $P_AA'=A$ and $t_q=1$, these maps satisfy $H_Xf_1=f_0\partial_1$, and $f_1(z\otimes t)=(z\otimes\varepsilon_j,0)$ for $z\in\ker A'$.

The LP structure and coordinate attachment also organize the physical implementation of the merged checks. In a reconfigurable atom array, each syndrome-extraction round is performed by shuttling syndrome ancillas to the data qubits in their check supports and applying entangling gates. The check weight fixes the number of ancilla--data gates, while the check connectivity determines the rearrangements needed between them. A crossed pair of orthogonal \textit{acousto-optic deflectors} (AODs) generates a two-dimensional array of movable optical tweezers. Selected atoms are picked up from static traps, transported by moving the AOD tweezers, and dropped off at their destinations. Within one such array, multiple rows or multiple columns can move in parallel, but their relative order along either axis must be preserved. We define the \textit{shuttling depth} $\delta_{\rm shuttle}$ as the number of sequential layers of collective AOD moves in one syndrome-extraction round. Each step of atom shuttling consists of picking up, moving, and dropping off atoms.

For unstructured checks, such as those of a generic graph-based surgery auxiliary, successive gate layers involve general ancilla--data matchings in Fig.~\ref{fig:overview}(b). Shuttling these matchings on a line of $w$ atoms uses general reshuffling, which admits $O(\log w)$ depth through parallel divide-and-conquer rearrangement~\cite{Xu2024ConstantOverhead}. Bounded check weight therefore limits the number of gate layers but does not by itself eliminate the shuttling overhead.

For the LP construction above, shuttling is organized at two scales. Let $n_a$ be the largest row or column dimension of the unlifted seed matrices $A,B,A',D$. At the \emph{protograph} level, ancillas are routed between the cyclic blocks specified by the unlifted seed matrices, requiring $O(\log n_a)$ depth for general matchings on at most $n_a$ seed coordinates. Each seed operation is repeated in parallel along the other LP coordinate as in Fig.~\ref{fig:overview}(a). At the \emph{lift} level, each monomial $x^s$ specifies the uniform cyclic shift $a\mapsto a+s\pmod l$, implemented in a constant number of collective moves as in Fig.~\ref{fig:overview}(c).

Combining these two levels gives three cases, assuming bounded total monomial weight per seed row and column and the layout and control conditions in the Supplemental Material (SM)~\cite{SM}.
\emph{(i)} For general seeds, routing between fibres takes $O(\log n_a)$ depth, followed by $O(\log l)$ depth to implement their different cyclic shifts for different $x^s$. These successive steps give $O(\log n_a+\log l)$ depth, matching general reshuffling of $l n_a$ atoms. Even with the same depth scaling, a compact two-dimensional layout can shorten travel distances relative to a one-dimensional register and reduce shuttling time.
\emph{(ii)} If only a bounded number of distinct monomials occur in the seed matrices, fibres requiring the same cyclic shift move together. The lift-level depth is then $O(1)$, leaving $O(\log n_a)$ for protograph routing.
\emph{(iii)} If the nonzero entries also lie on a bounded number of diagonals, connections along each diagonal have the same displacement, up to cyclic wraparound. Protograph routing then also takes $O(1)$ depth.
All three bounds apply to memory and surgery, including aligned attachment in Eq.~\eqref{eq:main-attachment} and return paths.

\begin{figure}[tb]
  \centering
  \includegraphics[width=\linewidth]{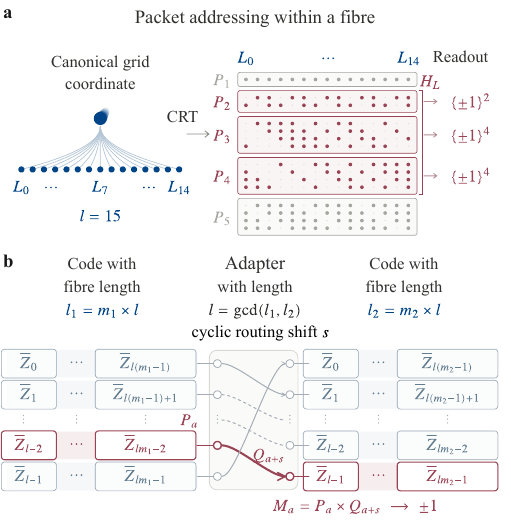}
  \caption{\textbf{Packet addressing and joint measurements between LP codes.}
    (a) A canonical logical fibre with $l=15$ has cyclic operators $L_0,\ldots,L_{14}$ and five CRT packets of dimensions $1,2,4,4,4$. Each dot row specifies a PPM. For instance, selecting $P_2$ gives 2 PPMs with binary outcomes. These packets are cyclically invariant subspaces that support overlapping logical PPMs.
  (b) An algebraic adapter can extract joint PPMs between logical fibres. For fibre lengths $l_1=m_1l$ and $l_2=m_2l$, with $l=\gcd(l_1,l_2)$, the adapter pairs row products $P_a$ and $Q_{a+s}$ and measures $M_a=P_aQ_{a+s}$, with indices understood modulo $l$. }
  \label{fig:packet-adapter}
\end{figure}

\paragraph{Addressability of algebraic surgery.}
\label{sec:addressibility}
The LP code has several useful logical bases and therefore supports a wide range of parallel measurement operations subject to cyclic symmetry constraints, as detailed in the SM~\cite{SM}. Here, as in hypergraph-product codes, the LP code over $R_l$ with odd $l$ organizes canonical logical operators in a grid by repeating kernel directions inherited from one seed code across rows supplied by the other seed code.
In the sector considered here, each column labels a canonical kernel direction of $A$, while each row labels a cokernel coordinate of $B$.
Each grid position carries a cyclic logical fibre of length $l$, whose operators are related by shifts of physical support [Fig.~\ref{fig:overview}(a)].
The K\"unneth formula~\cite{QGPU,Lee2026LogicalSpectroscopy,ZhengZhengJiangXu2026CanonicalLP} makes this picture precise through $H_1(C)\simeq(\ker A\otimes\operatorname{coker} B)\oplus(\operatorname{coker} A\otimes\ker B)$, where $C$ is the associated chain complex and the grid above describes canonical directions in the first summand.
For odd $l$, $x^l+1=\prod_c g_c$ has distinct irreducible factors, and the \emph{Chinese remainder theorem}
(CRT) gives $R_l\simeq\prod_c E_c$ with $E_c=\mathbb F_2[x]/(g_c)$.
This decomposition resolves each fibre into translation-invariant packets along its one-dimensional cyclic direction [Fig.~\ref{fig:packet-adapter}(a)].
Lemma~\ref{lem:two-bases} states the corresponding cyclic and CRT-adapted bases, with details in the SM~\cite{SM}.

\begin{lemma}[Cyclic and CRT-adapted logical bases]
  \label{lem:two-bases}
  Let $J$ and $J^\dagger$ be coordinate sets contained in the chosen information sets for $\ker A$ and $\operatorname{coker} B$, respectively, in every CRT component. There exist normalized kernel vectors $u_i\in\ker A$ and coordinate representatives $\varepsilon_j$ defining
  a canonical logical submodule $H_{\rm can}^{(L)}
  =\bigoplus_{i\in J,\,j\in J^\dagger}R_lq_{i,j}$ with $q_{i,j}=[(u_i\otimes\varepsilon_j,0)]$.
  Writing $\epsilon_c$ for the CRT idempotents and $d_c=\deg g_c$, each fibre $R_lq_{i,j}$ admits a cyclic and a CRT-adapted binary basis
  \begin{equation}
    \{x^a q_{i,j}\}_{a=0}^{l-1},
    \qquad
    \{\epsilon_c x^b q_{i,j}\}_{c,\,0\le b<d_c}.
    \label{eq:main-two-bases}
  \end{equation}
  The CRT-adapted basis expands in the cyclic basis through the invertible binary matrix $ T_{(c,b),a}
  =[x^a]\bigl(\epsilon_c x^b\bmod(x^l+1)\bigr)$.
\end{lemma}
These two bases expose two levels of addressability within the canonical logical submodule: a full \emph{logical fibre} and a \emph{logical packet}, which is a subspace of that fibre invariant under cyclic translation (Fig.~\ref{fig:packet-adapter}(a)). All these constructions are carried out over the polynomial ring, ensuring efficient AOD shuttling.

To address these logical submodules, we adapt the product-code modification approach of Refs.~\cite{ZhengJiangXu2025HighRateSurgery,ZhengZhengJiangXu2026CanonicalLP} to LP codes.
We append a logical-selection matrix $H_L$ to $A$, giving $A'=
\begin{pmatrix}A\\H_L
\end{pmatrix}$ and
$\ker A'=\ker A\cap\ker H_L$.
For a target fibre $R_lq_{i,j}$, the selection matrix acts on the kernel coordinate $i$, while the attachment below fixes the cokernel coordinate $j$. Let $\varepsilon_k\in R_l^{n_A}$ denote the standard coordinate vectors. The canonical kernel representatives satisfy
$\varepsilon_k^{\mathsf T}u_h=\delta_{k,h}$ for $k,h\in J$.
Writing
$J_i=J\setminus\{i\}$ and
$\overline J_i=\{0,\ldots,n_A-1\}\setminus J_i$,
the two selection matrices are
\begin{equation}
  H_{L,i}^{\mathrm{fib}}
  =\bigl(\varepsilon_k^{\mathsf T}\bigr)_{k\in J_i},\
  H_{L,i,S}^{\mathrm{CRT}}
  =
  \begin{pmatrix}
    \bigl(\varepsilon_k^{\mathsf T}\bigr)_{k\in J_i}\\
    \bigl(g_S \varepsilon_k^{\mathsf T}\bigr)_{k\in\overline J_i}
  \end{pmatrix},
  \label{eq:hl-resolution}
\end{equation}
where indexed rows are stacked vertically and
$g_S(x)=\prod_{c\in S}g_c(x)$ is interpreted as an element of $R_l$, where $S$ is the set of chosen CRT packets.
Within the canonical kernel span, the two matrices retain $R_lu_i$ and $\bigoplus_{c\in S}\epsilon_cR_lu_i$, respectively.
The packet restriction follows because $g_S$ vanishes in the selected CRT components and is invertible elsewhere.
These retained directions determine the canonical part of the measured subgroup in Eq.~\eqref{eq:css-surgery-readout} before binary expansion, as shown in the SM~\cite{SM}. In practice, we will choose some low-weight combination of $g_S$ to reduce the check overhead.

The same mechanism extends to addressing several fibres in one row or column, with a possibly different set of CRT packets selected on each fibre. Measuring more packets or fibres only changes $H_L$ while keeping $A$ and $D$ unchanged. The shared part of the auxiliary-code cost is therefore amortized over all retained PPMs. Explicit constructions, together with the resulting average auxiliary-code space cost per independent PPM, are given in the SM~\cite{SM}.

In addition, a similar construction, called the \emph{algebraic adapter}, couples two logical fibres and measures products of their operators, as illustrated in Fig.~\ref{fig:packet-adapter}(b).
For equal lifts $l_1=l_2=l$, one linking row joins the two selected factors according to
\begin{equation}
  A'_{\rm ad}=
  \begin{pmatrix}A'_1&0\\0&A'_2\\
    \varepsilon_{i_1}^{\mathsf T}&x^{-s} \varepsilon_{i_2}^{\mathsf T}
  \end{pmatrix},
  \label{eq:main-adapter}
\end{equation}
Here each $A'_\mu$ selects fibre $i_\mu$, and the linking row enforces $\eta_2=x^s\eta_1$ on their cyclic coefficients.
The adapter yields the $l$ products $P_aQ_{a+s}$, $a\in\mathbb Z_l$, as its canonical readout in one merged construction.
For unequal odd lifts with $e=\gcd(l_1,l_2)>1$, a common-period constraint yields $e$ canonical PPMs between matching residue-class products.
For example, lifts $33$ and $11$ pair each product of three translated operators in the first fibre with one operator in the second.
The unequal-lift construction and any surviving residual readouts are detailed in the SM~\cite{SM}.

\begin{figure}[t]
  \centering
  \includegraphics[width=\linewidth]{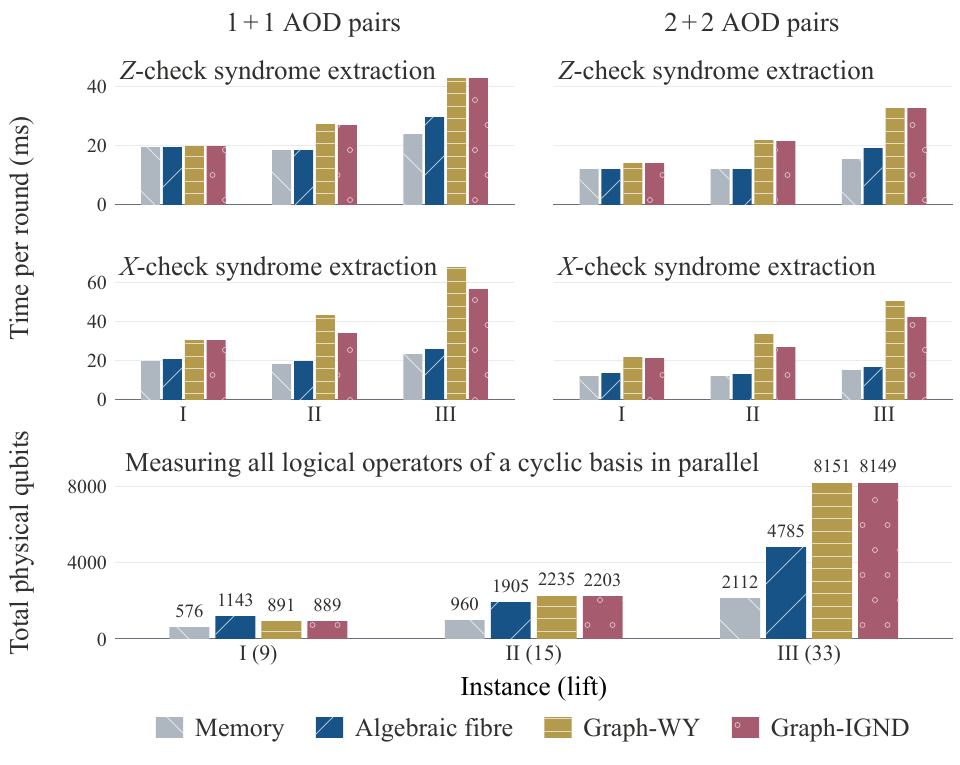}
  \caption{
    \textbf{Shuttling time and total space overhead.}
    \textit{Top two rows:} shuttling and gate times per $Z$- and $X$-check syndrome-extraction round, using one or two crossed AOD pairs per-code-per-basis. Memory uses only the data code.
    Instances I--III have lifts $l=9,15,33$, data codes $\code{306,52,8}$, $\code{510,76,10}$, $\code{1122,148,\leq20}$, and algebraic merged codes $\code{585,41,8}$, $\code{975,59,10}$, $\code{2442,113,\leq20}$, respectively.
    We denote the graph-based methods of Refs.~\cite{WilliamsonYoder2026,IdeGowda2025} by Graph-WY and Graph-IGND, respectively; their timings cover one observable on that fibre only.
    \textit{Bottom:} total physical qubits for parallel measurement of all logical operators of a cyclic basis, with memory as the baseline. Counts include all data and check qubits of the merged codes.
    Space counts for graph-based methods include $l$ disjoint surgery auxiliary qubits and shared data.
  }
  \label{fig:aod-timing}
\end{figure}

\paragraph{Numerical simulations.}
\label{sec:numerics}
To evaluate the time and physical-qubit overheads of surgery, we use three LP data codes, $\code{306,52,8}$, $\code{510,76,10}$, and $\code{1122,148,d\leq20}$, with lifts $l=9,15,33$, respectively. We provide a full animation of the shuttling simulation~\cite{repo_tgz}.
Fig.~\ref{fig:aod-timing} reports shuttling and gate times for one syndrome-extraction round.
In the algebraic layout, data and auxiliary atoms occupy separate two-dimensional arrays of physical fibres, each containing $l$ atoms in a smaller grid (Fig.~\ref{fig:overview}). The graph-based auxiliary occupies a one-dimensional register beside the data array.
The benchmark focuses on logical $Z$ measurements. The algebraic merged code measures all $l$ canonical logical $Z$ operators in the cyclic basis of one fibre, whereas each Graph-WY~\cite{WilliamsonYoder2026} or Graph-IGND~\cite{IdeGowda2025} gadget targets one fixed canonical logical $Z$ operator. The two graph-based algorithms are similar but differ slightly in how they initialize and grow the graph.
We take the atomic site spacing to be $12\,\mu\mathrm{m}$, the maximum acceleration to be $a_{\max}=5500\,\mathrm{m\,s^{-2}}$, and each entangling gate to last $1\,\mu\mathrm{s}$~\cite{LukinQEC,Xu2024ConstantOverhead,bluvstein2022quantumprocessora,zhao2026ultrahighratequantumerrorcorrection,BhardwajMaMeister2026}.
In the rest-to-rest transport model, a move of length $L$ accelerates and then decelerates at magnitude $a_{\max}$, taking $\tau(L)=2\sqrt{L/a_{\max}}$.
\begin{figure}[b]
  \centering
  \includegraphics[width=\linewidth]{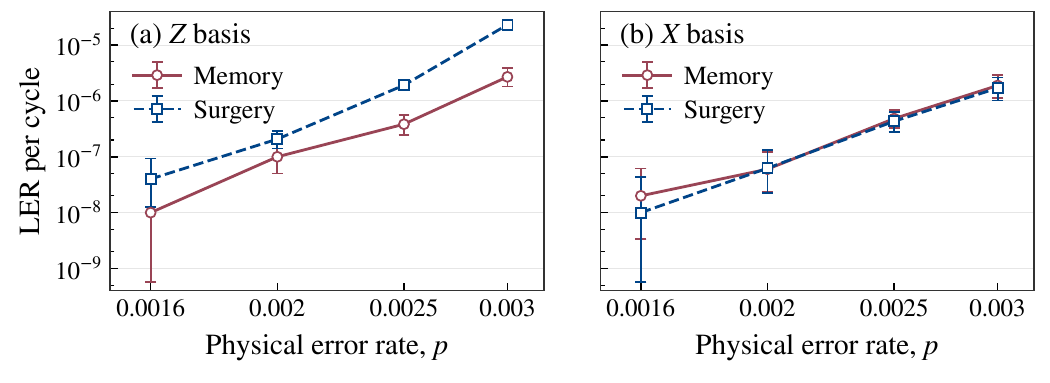}
  \caption{\textbf{Circuit-level logical error rates for $l=33$.}
  Memory and $X$-type algebraic fibre surgery are compared using \textbf{(a)} $Z$- and \textbf{(b)} $X$-basis preparation and readout. Memory tracks 148 logical observables. In (a), surgery tracks $148-(33+2)$ surviving logical observables, excluding two residual operators. In (b), surgery tracks 148 logical observables from the data-code readout and 33 from the fibre-surgery readout. Effective per-cycle rates are obtained from 20-round experiments decoded by Relay-BP with a BP--LSD fallback. Error bars show 95\% confidence intervals.}
  \label{fig:ler}
\end{figure}
The reported times include shuttling, docking, gates, and scheduled returns with the implemented inter-block passing lanes and return and interface detours, but exclude readout, reset, trap transfers, and within-block collision avoidance for the LP layout.
We assign one or two crossed AOD pairs to each basis of both the data and auxiliary codes, while memory uses only the data AOD pair.
At $l=33$, algebraic surgery increases the $Z$-check time relative to memory by $24\%$ with one AOD pair per-code-per-basis and $23\%$ with two, while either graph-based schedule increases it by about $81\%$ and $113\%$, respectively.
For $X$ checks, the algebraic overhead is $10\%$ with one pair and $9\%$ with two, compared with $142$--$191\%$ for the graph schedules with one pair and $180$--$233\%$ with two. Furthermore, with independent $X/Z$ AOD control, the recent left–right circuits method~\cite{StrikisBrowneBeverland2026} allows the $X$- and $Z$-check schedules to run in parallel~\cite{SM}.
An animation of the simulated shuttling sequences is available online~\cite{repo_tgz}.

For parallel measurement of all $l$ cyclic-basis logical $Z$ operators of one fibre, Fig.~\ref{fig:aod-timing} counts data and check qubits: algebraic surgery shares one auxiliary LP code, whereas the plotted Graph-WY and Graph-IGND constructions share the data code but replicate $l$ single-observable gadgets.
At $l=33$, this gives 4785 qubits for algebraic surgery versus 8151 and 8149 for the two graph constructions, about $41\%$ fewer qubits in this specified comparison.
A separate parallel hypergraph surgery construction~\cite{ZhengZhengJiangXu2026CanonicalLP} achieves competitive space overhead, but its surgery still breaks the regular 2D grid structure and could potentially incur higher time overhead, leaving a detailed comparison for future work.

We next compare circuit-level LERs for memory and $X$-type algebraic fibre surgery at $l=33$ in Fig.~\ref{fig:ler}.
In both readout bases, data are prepared and measured in the chosen basis, while both $X$ and $Z$ stabilizers of the data (memory) or merged (surgery) code are measured for 20 rounds using the same edge-colouring schedule.
Surgery also prepares the auxiliary qubits in $|0\rangle$ and measures them in $Z$ to detach them. Products of first-round added-stabilizer outcomes give the $X$-type surgery observables in the $X$-basis experiment, whereas the $Z$-basis experiment tracks surviving data logicals.
The retained data $X$ checks protect against $Z$ errors, while protection against $X$ errors also depends on the auxiliary factor $D$~\cite{SM}.
Gate depolarization and preparation and measurement flips each occur with probability $p$. Idling errors are not included.
Relay-BP~\cite{mullerImprovedBeliefPropagation2025b} decodes each shot first, with BP-LSD~\cite{hillmannLocalizedStatisticsDecoding2025} applied to non-convergent one~\cite{zhao2026ultrahighratequantumerrorcorrection}.
Over the sampled range, surgery has a memory-comparable $X$-basis LER and a higher $Z$-basis LER that remains within one order of magnitude of memory.
Simulation details are given in the SM~\cite{SM}.

\paragraph{Discussion.}
\label{sec:discussion}
Algebraic surgery uses the group-algebra structure of lifted-product
codes to co-design the measured logical operators, the auxiliary code, and
the hardware implementation. At the physical level, preserving the cyclic
structure of the LP code allows the relevant check interactions to be
organized into a bounded number of collective shuttling stages, giving
constant shuttling depth
under the stated layout assumptions. Our finite-size
shuttling simulations show that the resulting surgery rounds can have cycle
times close to those of the corresponding memory baselines, whereas the
graph-based schedules considered here require substantially longer cycles
even for a single logical observable. Graph-based constructions offer more
general logical addressability, whereas algebraic surgery trades some
generality for cyclic logical symmetry and structured parallel access.
Canonical operators are organized into cyclic fibres and CRT packets,
enabling parallel inter-fibre $XX$ and $ZZ$ measurements, including products
whose physical representatives overlap. At the
circuit level, the tested surgery has a memory-comparable $X$-basis LER
but a higher $Z$-basis LER.

A natural application of this structured logical symmetry is the simulation of lattice spin Hamiltonians. Logical fibres provide a natural way to realize one-dimensional cyclic chains of logical qubits, and interactions in a lattice model can be encoded as global couplings between such chains. The Clifford part can then be implemented through parallel inter-fibre PPMs with cyclic shifts.
Compared with the architectures in Ref.~\cite{ismail2026fastparallelstar,xu2025batchedhighrate}, which allow transversal implementation of such Clifford gates but require a disjoint logical basis, the LP construction considered here permits overlapping physical representatives and can provide a higher logical encoding efficiency. CRT-packet selection can therefore implement fibre-internal translation-invariant products, including the weight-2 intra-fibre PPMs that arise in next-nearest-neighbour interactions. Establishing a practical advantage will require detailed
resource estimates for algorithm compilation and fault-tolerant gadget
evaluation, including parallel magic state preparation and teleportation.

Several directions could extend this framework beyond cyclic lifted-product codes. First, applying the same design principles to non-product cyclic permutation codes, such as pair-partition codes~\cite{okada2026pairpartitionconstructionscpmbasedquantum,LeeOkadaMaskara2026}, could further improve the logical encoding rate while retaining cyclic logical symmetry~\cite{OkadaKasai2026LogicalBases} and structured shuttling. Second, non-Abelian group algebras may provide richer permutation symmetries, but would require a specialized addressing mechanism because the canonical basis and packet structure used here are not generally available. Recent non-Abelian lifted-product constructions, such as the
Mitten code~\cite{BhardwajMaMeister2026}, introduce additional conditions under which a canonical basis can still be constructed, providing a possible starting point for extending fibre-based logical operations beyond Abelian cyclic groups. More generally, cyclic symmetry is only one example of a useful
organizing principle: other translation, reflection, or permutation
symmetries could define different logical-fibre structures tailored to a
target application. Such symmetry-driven constructions could lead to
application-specific fault-tolerant architectures in which the code,
logical measurements, and hardware implementation are designed jointly.

\label{main:text-end}
\paragraph{Acknowledgements.}
C.Z. thanks Hengyun Zhou, Nishad Maskara, and Arpit Dua for helpful discussions. T.-G.Z. thanks Yifei Wang and Xixiang Du for helpful discussions. We thank Zi-Han Chen, Ying Li, Qian Xu, and Hengyun Zhou for their suggestions on the early version of this manuscript. T.-G.Z. was supported by the Swiss National Science Foundation under Division II (Grant No.~200020-219400). The Berlin team has been supported by the BMFTR (QSolid, MUNIQC-Atoms, PasQuops), the DFG (CRC 183, BoLaCo, and SPP 2514), the Quantum Flagship (Millenion, PasQuanS2), the Munich Quantum Valley, Berlin Quantum, QuantERA (SDPCode), and the European Research Council (DebuQC). Part of the computational work for this research was supported by the High Performance Computing Center at Eastern Institute of Technology, Ningbo. The authors acknowledge the use of OpenAI's GPT-6 and Moonshot AI's Kimi-K3 for visualizing atom-shuttling animations, running numerical simulations, performing theoretical derivations, and preparing the manuscript. The authors take full responsibility for the results.

\paragraph{Note.}
During the preparation of this manuscript, we became aware of two related independent works~\cite{LiftedSurgery,MaskaraEfficientLogic2026}. The lifted-surgery construction of Ref.~\cite{LiftedSurgery} provides single-shot surgery systems that preserve the algebraic structure of the underlying code, while Ref.~\cite{MaskaraEfficientLogic2026} develops symmetry-based logical operations for ultra-high-rate quantum codes with low space-time overheads. Our results are complementary: we provide a concrete construction based on LP codes and focus on hardware efficiency for neutral-atom processors, supported by detailed shuttling simulations.

\clearpage
\bibliographystyle{apsrev4-2}
\bibliography{references}

\clearpage
\onecolumngrid
\setcounter{page}{1}
\setcounter{equation}{0}\setcounter{figure}{0}\setcounter{table}{0}\setcounter{subsection}{0}
\renewcommand{\thefigure}{S\arabic{figure}}
\renewcommand{\thetable}{S\arabic{table}}
\renewcommand{\theequation}{S\arabic{equation}}
\renewcommand{\thesubsection}{S\Roman{subsection}}
\section*{Supplemental Material for\\ Parallel algebraic surgery for qLDPC codes with constant shuttling depth}
\label{supp:start}
This supplement contains the extended constructions, proofs, shuttling records, and numerical examples accompanying the Letter. References and the main-text equation numbers refer to the accompanying Letter.

\section{Algebraic surgery method}

\subsection{CSS surgery as a mapping cone}

A CSS data code is a length-2 chain complex with parity-check matrices $H_X,H_Z$ satisfying $H_XH_Z^\dagger=0$. The auxiliary system is another length-2 complex $\mathcal A_1\xrightarrow{\partial_1}\mathcal A_0\xrightarrow{\partial_0}\mathcal A_{-1}$, with auxiliary qubits in $\mathcal A_0$ and $\partial_0\partial_1=0$. Attachment maps $f_1$ and $f_0$ satisfying $H_Xf_1=f_0\partial_1$ connect the auxiliary and data complexes~\cite{IdeGowda2025}
according to
\begin{equation}
  \begin{tikzcd}[column sep=large,row sep=large]
    \mathcal{A}_1 \arrow[r,"\partial_1"] \arrow[dr,"f_1"'] &
    \mathcal{A}_0 \arrow[r,"\partial_0"] \arrow[dr,"f_0"] &
    \mathcal{A}_{-1} \\
    C_2 \arrow[r,"H_Z^\dagger"'] &
    C_1 \arrow[r,"H_X"'] &
    C_0
  \end{tikzcd}.
  \label{eq:css-surgery-diagram}
\end{equation}
Their total complex,
$\mathcal A_1\oplus C_2\xrightarrow{\widetilde H_Z^\dagger}
\mathcal A_0\oplus C_1\xrightarrow{\widetilde H_X}
\mathcal A_{-1}\oplus C_0$,
defines the merged code with checks
\begin{equation}
  \widetilde H_Z=
  \begin{pmatrix}
    H_Z&0\\
    f_1^\dagger&\partial_1^\dagger
  \end{pmatrix},
  \qquad
  \widetilde H_X=
  \begin{pmatrix}
    H_X&f_0\\
    0&\partial_0
  \end{pmatrix}.
  \label{eq:sm-css-surgery-checks}
\end{equation}
Here $H^\dagger$ denotes binary transpose, or transpose combined with $x\mapsto x^{-1}$ over the cyclic ring $\mathbb F_2[C_l]\cong\mathbb F_2[x]/(x^l+1)$~\cite{9567703,QGPU}.

\subsection{Extended LP construction and logical bases}
\label{em:lp-construction}

\paragraph{LP complex and binary checks.}
Let $R_l=\mathbb F_2[x]/(x^l+1)$ with odd $l$, and take all tensor products over $R_l$ unless stated otherwise. For factor maps $A:A_1=R_l^{n_A}\to A_0=R_l^{m_A}$ and $B:B_1=R_l^{n_B}\to B_0=R_l^{m_B}$, the LP complex is obtained by totalizing
\begin{equation}
  \begin{gathered}
    \begin{tikzcd}[
        ampersand replacement=\&,
        baseline=(current bounding box.center),
        row sep=2.2em
      ]
      C_0 \\
      C_1 \arrow[u,"H_X"] \\
      C_2 \arrow[u,"H_Z^\dagger"]
    \end{tikzcd}
    \qquad
    \begin{tikzcd}[
        ampersand replacement=\&,
        baseline=(current bounding box.center),
        row sep=2.2em,
        column sep=1.6em
      ]
      \& A_0\otimes B_0 \& \\
      A_1\otimes B_0 \arrow[ur,"A\otimes I"]
      \& \& A_0\otimes B_1 \arrow[ul,swap,"I\otimes B"] \\
      \& A_1\otimes B_1
      \arrow[ul,"I\otimes B"]
      \arrow[ur,swap,"A\otimes I"] \&
    \end{tikzcd}
  \end{gathered}
  \label{eq:lp-complex}
\end{equation}
with $C_1=(A_1\otimes B_0)\oplus(A_0\otimes B_1)$, $C_0=A_0\otimes B_0$, and $C_2=A_1\otimes B_1$. The adjoint $\dagger$ combines matrix transposition with $x\mapsto x^{-1}$. In the displayed ordering,
\begin{equation}
  H_X=\bigl(A\otimes I_{m_B}\mid I_{m_A}\otimes B\bigr),
  \qquad
  H_Z^\dagger=
  \begin{pmatrix}
    I_{n_A}\otimes B\\
    A\otimes I_{n_B}
  \end{pmatrix},
  \label{eq:lp-checks}
\end{equation}
and the two paths through the product complex cancel in characteristic two,
$H_XH_Z^\dagger=A\otimes B+A\otimes B=0$.
Binarization replaces $x^s$ by the circulant $P(s)$ with $P(s)_{a,a+s}=1$ modulo $l$. It preserves products and sends $\dagger$ to binary transpose, so $\widehat H_X\widehat H_Z^{\mathsf T}=0$.

\paragraph{Canonical logical submodule.}
Since $l$ is odd, $x^l+1=\prod_c g_c$ is square-free and the Chinese remainder theorem (CRT) gives $R_l\simeq\prod_cE_c$, where $E_c=\mathbb F_2[x]/(g_c)$. The K\"unneth decomposition contains no torsion term,
\begin{equation}
  H_1(C)\simeq H_1^{(L)}\oplus H_1^{(R)},\qquad
  H_1^{(L)}=\ker A\otimes\operatorname{coker}B,\qquad
  H_1^{(R)}=\operatorname{coker}A\otimes\ker B .
  \label{eq:kunneth}
\end{equation}
We describe the left sector. The right follows by exchanging the factors.

In CRT component $c$, set $A_c=A\bmod g_c$ and $B_c=B\bmod g_c$. With fixed coordinate orderings, let $J_c$ and $J_c^\dagger$ be the nonpivot columns of the reduced row-echelon forms $\operatorname{RREF}(A_c)$ and $\operatorname{RREF}(B_c^{\mathsf T})$, respectively. Choose systematic kernel vectors $u_i^{(c)}$ satisfying $(u_i^{(c)})_{i'}=\delta_{ii'}$ for $i,i'\in J_c$. The coordinate vectors $\varepsilon_j$, $j\in J_c^\dagger$, represent a basis of $\operatorname{coker}B_c$, giving packet logical classes $q_{i,j}^{(c)}=u_i^{(c)}\otimes_{E_c}[\varepsilon_j]$. When block labels are required, we write $u_i^{[\mu],(c)}$.

The common sets $J=\bigcap_cJ_c$ and $J^\dagger=\bigcap_cJ_c^\dagger$ define the canonical submodule. Let $\epsilon_c$ be the CRT idempotents, $\epsilon_c\bmod g_d=\delta_{cd}$, and define $u_i=\sum_c\epsilon_cu_i^{(c)}\in\ker A$. For $i\in J$ and $j\in J^\dagger$, set $q_{i,j}=[(u_i\otimes\varepsilon_j,0)]$. Then
\begin{equation}
  H_{\mathrm{can}}^{(L)}
  =\bigoplus_{i\in J,\,j\in J^\dagger}R_lq_{i,j}
  \simeq
  \bigoplus_c\bigoplus_{i\in J,\,j\in J^\dagger}E_cq_{i,j}^{(c)} .
  \label{eq:logical-fibres}
\end{equation}
The remaining packet directions form the complementary residual submodule of $H_1^{(L)}$.

\paragraph{Two logical bases.}
Each canonical fibre has a cyclic-orbit basis and a CRT power basis,
\begin{equation}
  \bar q_{i,j;a}^{\mathrm{cyc}}=x^aq_{i,j},
  \qquad
  \bar q_{i,j;b}^{\mathrm{CRT}(c)}=\epsilon_cx^bq_{i,j},
  \qquad
  a\in\mathbb Z_l,\quad 0\le b<d_c=\deg g_c .
  \label{eq:packet-basis}
\end{equation}
The former makes cyclic translation explicit, whereas the latter resolves the invariant CRT packets. Writing $\epsilon_cx^b=\sum_{a=0}^{l-1}T_{c,b,a}x^a$ gives
\begin{equation}
  \bar q_{i,j;b}^{\mathrm{CRT}(c)}
  =\sum_{a=0}^{l-1}T_{c,b,a}\bar q_{i,j;a}^{\mathrm{cyc}},
  \qquad T_{c,b,a}\in\mathbb F_2,
\end{equation}
so a CRT basis vector generally represents a product of cyclic logical operators rather than a single cyclic operator.

\paragraph{Proof of Lemma~\ref{lem:two-bases}.}
The packet kernel and cokernel bases have independent tensor products. Any relation among the canonical generators therefore has coefficients that vanish modulo every $g_c$, and hence vanish in $R_l$, proving that each $R_lq_{i,j}$ is a free rank-one summand. Transporting the two binary bases $\{x^a\}_{a=0}^{l-1}$ and $\{\epsilon_cx^b:0\le b<d_c\}_c$ of $R_l$ to this summand proves Eq.~\eqref{eq:main-two-bases}. Since $\sum_cd_c=l$, the change-of-basis matrix $T$ is invertible. For conjugate logical Pauli bases, a $Z$-basis transformation $T$ requires the dual $X$-basis transformation $T^{-\mathsf T}$.

\subsection{Auxiliary LP code construction}
\label{sm:selection-attachment}

We first establish the auxiliary code and attachment for a general selected factor $A':R_l^{n_A}\to R_l^{m_A+s}$ satisfying $P_AA'=A$, where $P_A=(I_{m_A}\mid0)$ projects onto the original rows of $A$. The next subsection constructs $A'$ for fibre and CRT-packet selection. Let $D:R_l^{n_D}\to R_l^{m_D}$ be a second factor. The auxiliary system is the LP code $\mathrm{LP}(A',D)$, $\mathcal A_1\xrightarrow{\partial_1}\mathcal A_0\xrightarrow{\partial_0}\mathcal A_{-1}$ with
\begin{equation}
  \partial_1=
  \begin{pmatrix}
    A'\otimes I_{n_D}\\
    I_{n_A}\otimes D
  \end{pmatrix},
  \qquad
  \partial_0=
  \bigl(I_{m_A+s}\otimes D\mid A'\otimes I_{m_D}\bigr).
  \label{eq:tube}
\end{equation}
We require the tube factor to have a rank-one $R_l$ kernel, $\ker D=R_lt$, and choose a coordinate $q$ for which $t_q$ is a unit. After rescaling, take $t_q=1$. For odd $l$, the CRT decomposition reduces the kernel calculation to vector spaces over the component fields. In each component,
$\ker(A'_c\otimes I)\cap\ker(I\otimes D_c)=\ker A'_c\otimes\ker D_c$,
and hence
\begin{equation}
  \ker\partial_1
  =\ker A'\otimes_{R_l}\ker D
  =\{z\otimes t:z\in\ker A'\}.
  \label{eq:sm-tube-kernel}
\end{equation}
Evaluation at coordinate $q$ recovers $z$, so thickening preserves the complete selected kernel of $A'$.

Besides this kernel condition, we choose $D$ to retain the hardware structure of the LP construction. In particular, monomial entries are directly compatible with cyclic AOD translations, while zeros can be used to reduce check weight and gate conflicts. In the examples below we choose sparse tube matrices with their zero pattern arranged so that several check interactions can be executed in parallel. These hardware considerations are separate from the algebraic requirement $\ker D=R_lt$.

To attach the auxiliary kernel to the data code $C=\mathrm{LP}(A,B)$, choose a cokernel coordinate $j$ in the $B_0$ factor and set $W=\varepsilon_j\varepsilon_q^{\mathsf T}$, where $\varepsilon_j\in R_l^{m_B}$ and $\varepsilon_q\in R_l^{n_D}$ are standard coordinate vectors. The attachment maps are
\begin{equation}
  f_1=
  \begin{pmatrix}
    I_{n_A}\otimes W\\
    0
  \end{pmatrix},
  \qquad
  f_0=\bigl(P_A\otimes W\mid0\bigr),
  \label{eq:em-tube-wiring}
\end{equation}
and $P_AA'=A$ gives $H_Xf_1=A\otimes W=f_0\partial_1$. For $\xi=z\otimes t\in\ker\partial_1$, the normalization $t_q=1$ further gives $f_1\xi=(z\otimes\varepsilon_j,0)$. The complete logical subspace measured by the surgery is therefore
\begin{equation}
  M_j(A')
  =
  \left\{
    [(z\otimes\varepsilon_j,0)]:
    z\in\ker A'
  \right\}
  \subseteq H_1(C).
  \label{eq:sm-measured-space}
\end{equation}
Thus $A'$ determines the measured kernel directions, $D$ supplies auxiliary redundancy without changing that kernel, and the coordinate attachment fixes their cokernel address.

This argument identifies the
measured logical space but does not establish a lower bound on the merged-code distance, which is assessed separately
for the finite constructions considered here. The tube also contributes to the hardware cost: a monomial entry of $D$ lifts to one cyclic matching, whereas an entry of polynomial weight $w$ requires up to $w$ cyclic translations. Sparse choices of $A'$, $D$, and the attachment therefore retain the structured routing
used by algebraic surgery.

\subsection{Fibre and CRT-packet selection}
\label{sm:selection}

We now construct $A'$ so that Eq.~\eqref{eq:sm-measured-space} addresses a prescribed canonical logical fibre or selected CRT packets within it. Let $J$ be the common aligned coordinate set for the normalized generators $u_i\in\ker A$, satisfying $\varepsilon_r^{\mathsf T}u_i=\delta_{r,i}$ for $r,i\in J$. For a target fibre $i$, define $J_i=J\setminus\{i\}$ and $\overline J_i=\{0,\ldots,n_A-1\}\setminus J_i$. For a set $S$ of CRT components, write $g_S=\prod_{c\in S}g_c$ and $e_S=\sum_{c\in S}\epsilon_c$ in $R_l$. The fibre and CRT selectors are
\begin{equation}
  H_{L,i}^{\mathrm{fib}}
  =\left(\varepsilon_r^{\mathsf T}\right)_{r\in J_i},
  \qquad
  H_{L,i,S}^{\mathrm{CRT}}
  =
  \begin{pmatrix}
    \left(\varepsilon_r^{\mathsf T}\right)_{r\in J_i}\\
    g_S\left(\varepsilon_r^{\mathsf T}\right)_{r\in\overline J_i}
  \end{pmatrix},
  \qquad
  A'=
  \begin{pmatrix}A\\H_L
  \end{pmatrix},
  \quad
  \ker A'=\ker A\cap\ker H_L.
  \label{eq:em-hl-resolution}
\end{equation}
The indexed coordinate rows are stacked vertically.

To determine the complete selected kernel, define
$K_{\mathrm{res}}=\{z\in\ker A:\varepsilon_r^{\mathsf T}z=0\text{ for every }r\in J\}$.
Systematic normalization gives
\begin{equation}
  \ker A
  =\bigoplus_{r\in J}R_lu_r\oplus K_{\mathrm{res}}.
  \label{eq:sm-kernel-decomposition}
\end{equation}
The fibre selector removes all canonical summands except $R_lu_i$ and leaves $K_{\mathrm{res}}$ unchanged. For CRT selection, $g_S$ vanishes in each component $c\in S$ and is invertible in every component $c\notin S$, restricting all surviving directions to $R_le_S$. Hence
\begin{equation}
  \ker A'_{\mathrm{fib}}=R_lu_i\oplus K_{\mathrm{res}},
  \qquad
  \ker A'_{\mathrm{CRT}}=R_l(e_Su_i)\oplus e_SK_{\mathrm{res}}.
  \label{eq:fibre-packet}
\end{equation}
The fibre selector therefore isolates the chosen generator within the canonical submodule, while CRT selection restricts it further to the chosen invariant packets. Taking all CRT components gives $e_S=1$ and recovers the full-fibre construction.

Combining Eq.~\eqref{eq:fibre-packet} with Eq.~\eqref{eq:sm-measured-space}, the canonical measured space is
\begin{equation}
  \mathcal W_{i,j,S}^{\mathrm{can}}
  =R_l\bigl[(e_Su_i\otimes\varepsilon_j,0)\bigr]
  =\bigoplus_{c\in S}\epsilon_cR_lq_{i,j},
  \qquad
  d_S=\dim_{\mathbb F_2}\mathcal W_{i,j,S}^{\mathrm{can}}
  =\sum_{c\in S}\deg g_c.
  \label{eq:sm-canonical-readout}
\end{equation}
For a chosen CRT power basis, the kernel vectors
\begin{equation}
  \xi_b^{(c)}=(\epsilon_cx^bu_i)\otimes t,
  \qquad c\in S,\quad 0\le b<\deg g_c,
\end{equation}
satisfy $\partial_1\xi_b^{(c)}=0$ and
$f_1\xi_b^{(c)}=(\epsilon_cx^bu_i\otimes\varepsilon_j,0)$.
Thus each CRT-adapted logical operator is available as a parity of the added-check outcomes. The carving fixes the measured subspace, but not a preferred ordered binary basis within it.

Coordinate selection can also retain residual logical directions. When $e_SK_{\mathrm{res}}=0$, the canonical packet is the complete readout in this sector. Otherwise, any nontrivial logical images of $e_SK_{\mathrm{res}}$ are measured in addition and must either be retained as part of the intended PPM or removed by further selection rows. The selector should therefore not be identified with the canonical fibre alone unless these residual contributions have been checked.

The routing cost of packet selection is controlled by the polynomial structure of $g_S$, rather than by $d_S$ alone. Several monomials require several cyclic translations after lifting, whereas unions of CRT packets can exhibit cancellations and remain sparse even when the selected subspace is large. For every proper divisor $q\mid l$, for example,
\begin{equation}
  \frac{x^l+1}{x^q+1}
  =\sum_{a=0}^{l/q-1}x^{aq}
\end{equation}
is a sparse divisor corresponding to a union of CRT components. Such sparse unions make selective packet surgery compatible with the same structured shuttling principles as full-fibre surgery. Table~\ref{tab:low-weight-unions} illustrates this effect for $l=33$.

\begin{table}[tb]
  \centering
  \caption{\textbf{Selected CRT unions for $l=33$.}
  Here $d_S=\dim_{\mathbb F_2}(R_le_S)=\sum_{c\in S}\deg g_c$ is the selected binary dimension within one canonical fibre. The final column gives the polynomial weight of $g_S$, which controls the number of cyclic translations associated with each selector cell, excluding additional rows used to remove residual directions. Low-weight choices are highlighted in bold.}
  \label{tab:low-weight-unions}
  \small
  \setlength{\tabcolsep}{4pt}
  \begin{tabular}{lclc}
    \toprule
    $S$ & $d_S$ & $g_S(x)$ & $\operatorname{wt}(g_S)$\\
    \midrule
    $\{0\}$ & $1$ & $1+x$ & $\mathbf{2}$\\
    $\{1\}$ & $2$ & $1+x+x^2$ & $\mathbf{3}$\\
    $\{2\}$ & $10$ & $1+x^3+x^5+x^7+x^{10}$ & $5$\\
    $\{3\}$ & $10$ & $1+x+x^5+x^9+x^{10}$ & $5$\\
    $\{4\}$ & $10$ & $1+x+\cdots+x^{10}$ & $11$\\
    \midrule
    $\{0,1\}$ & $\mathbf{3}$ & $1+x^3$ & $\mathbf{2}$\\
    $\{0,4\}$ & $\mathbf{11}$ & $1+x^{11}$ & $\mathbf{2}$\\
    $\{1,2,3\}$ & $\mathbf{22}$ & $1+x^{11}+x^{22}$ & $\mathbf{3}$\\
    \bottomrule
  \end{tabular}
\end{table}

\subsection{Parallel measurements with one auxiliary code}
\label{sm:high-rate}

Several fibres or packets in the same row of the canonical grid can be measured with one auxiliary code. They share the cokernel coordinate $j$ and the attachment in Eq.~\eqref{eq:em-tube-wiring}.
Let $\mathcal C$ label the CRT factors in $x^l+1=\prod_{c\in\mathcal C}g_c$. The set $\mathcal T\subseteq J\times\mathcal C$ lists the packets chosen from each fibre. For fibre $i$, the chosen packets are $S_i=\{c:(i,c)\in\mathcal T\}$. Different fibres can use different packet sets. The selected part of the canonical kernel is
\begin{equation}
  K_{\mathcal T}^{\rm can}
  =\bigoplus_{(i,c)\in\mathcal T}\epsilon_cR_lu_i.
  \label{eq:high-rate-target-kernel}
\end{equation}
For each $i\in J$, write $g_{S_i}=\prod_{c\in S_i}g_c$, with $g_{\varnothing}=1$. We select these packets by adding the following rows to $A$. When $S_i=\mathcal C$, the row is zero in $R_l$ and can be omitted.
\begin{equation}
  H_{L,\mathcal T}^{\rm can}
  =\left(g_{S_i}\varepsilon_i^{\mathsf T}\right)_{
  i\in J:\,S_i\ne\mathcal C},
  \qquad
  \ker\!
  \begin{pmatrix}A\\H_{L,\mathcal T}^{\rm can}
  \end{pmatrix}
  =K_{\mathcal T}^{\rm can}\oplus K_{\rm res}.
  \label{eq:high-rate-general-selector}
\end{equation}
In CRT component $c$, the row $g_{S_i}\varepsilon_i^{\mathsf T}$ forces the coefficient of $u_i$ to zero when $c\notin S_i$ and leaves it free when $c\in S_i$. Extra selector rows can remove unwanted directions in $K_{\rm res}$. We call the selector, including any such rows, $H_{L,\mathcal T}$.

We then form the auxiliary code with the same $D$ and attach it at $j$. By Eq.~\eqref{eq:sm-measured-space}, the canonical logical operators measured span the space
\begin{equation}
  \mathcal W_{\mathcal T,j}^{\rm can}
  =\bigoplus_{(i,c)\in\mathcal T}\epsilon_cR_lq_{i,j}.
  \label{eq:high-rate-row-readout}
\end{equation}
Each selected PPM is obtained from a parity of check outcomes in this merged code. These PPMs are independent and share the same auxiliary code. As in the single-fibre case, residual kernel directions must be removed by extra selector rows or included in the intended measurements.

To measure fibres or packets in one column, exchange the two factors in the LP complex and the $X$ and $Z$ roles. Then repeat Eqs.~\eqref{eq:high-rate-target-kernel}--\eqref{eq:high-rate-row-readout} with the selector and attachment for the exchanged factors. Simply attaching one fixed rank-one auxiliary kernel to several cokernel coordinates would measure their product. It would not give independent outcomes for each coordinate.

\begin{figure}[tb]
  \centering
  \begin{tikzpicture}
    \node[inner sep=0] (highrateplot) {\includegraphics[width=0.92\linewidth]{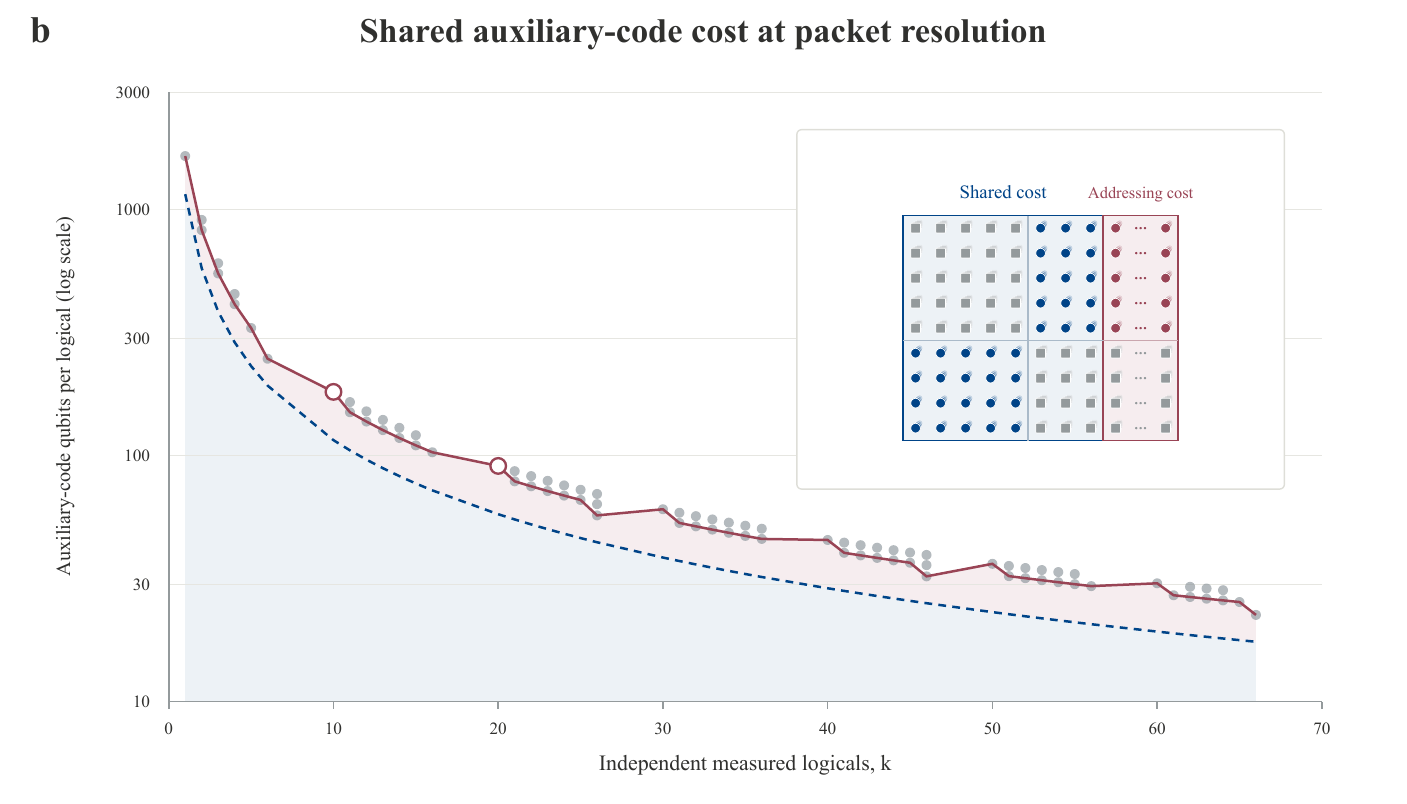}};
    \fill[white] (highrateplot.north west) rectangle ([xshift=0.06\linewidth,yshift=-0.06\linewidth]highrateplot.north west);
  \end{tikzpicture}
  \caption{\textbf{Auxiliary qubits per parallel measurement.}
  We consider the $l=33$ instance with two canonical fibres in one row. Each fibre has packets of dimensions $1,2,10,10,10$. Gray points show the number of auxiliary-code qubits per independent PPM for all $1023$ nonempty packet selections. The solid curve gives the lowest cost for each number $k$ of independent PPMs. The dashed curve shows the shared cost $N_{\rm shared}/k$. The gap between the curves is the cost $N_{\rm address}/k$ of the selector rows for the lowest-cost selection. This qubit count does not establish the distance of the merged code.}
  \label{fig:high-rate-space-cost}
\end{figure}

If $H_{L,\mathcal T}$ has $s_{\mathcal T}$ rows over $R_l$, the number of qubits in the auxiliary code $\mathrm{LP}(A'_{\mathcal T},D)$ is
\begin{align}
  N_{\rm aux}(\mathcal T)
  &=l\bigl[(m_A+s_{\mathcal T})n_D+n_Am_D\bigr]\nonumber\\
  &=\underbrace{l(m_An_D+n_Am_D)}_{N_{\rm shared}}
  +\underbrace{l s_{\mathcal T}n_D}_{N_{\rm address}}.
  \label{eq:high-rate-space-cost}
\end{align}
Here $N_{\rm shared}$ is the number of qubits shared by all selected measurements, and $N_{\rm address}$ counts the qubits added by the selector rows. The chosen packets give $k_{\mathcal T}=\sum_{(i,c)\in\mathcal T}\deg g_c$ independent canonical PPMs, so the average number of auxiliary qubits per PPM is
$ \bar n_{\rm aux}(\mathcal T)
=\frac{N_{\rm shared}}{k_{\mathcal T}}
+\frac{N_{\rm address}}{k_{\mathcal T}}$.
Fig.~\ref{fig:high-rate-space-cost} shows this saving for the $l=33$ instance with two fibres. Measuring more PPMs together generally lowers the number of auxiliary qubits per PPM because the shared cost is spread over more measurements. The spread of points at a fixed $k$ comes from packet selections that require different numbers of selector rows.
This count excludes data qubits and ancillas used to measure syndromes. It gives the number of auxiliary-code qubits needed when all selected PPMs are measured at once. Reusing auxiliary qubits for measurements in sequence can reduce this number, but takes more time.

\subsection{Algebraic adapters}
\label{sm:adapter}

The preceding construction selects and measures a logical submodule of a single LP block. An \emph{algebraic adapter} instead couples selected kernel directions from two data blocks before applying the same tube and coordinate-attachment construction. We write $Q_\mu$, $\mu=1,2$, for the two blocks, use superscripts $[\mu]$ for their logical generators, and denote the selected fibre coordinates by $(i_\mu,j_\mu)$.

\paragraph{Equal lifts.}
Suppose both blocks are defined over $R_l=\mathbb F_2[x]/(x^l+1)$. Let $A'_\mu$ denote the factor obtained after selecting fibre $i_\mu$, or the same CRT packet set $S$, in block $\mu$. Equation~\eqref{eq:fibre-packet} gives
\begin{equation}
  \ker A'_\mu
  =R_le_Su_{i_\mu}^{[\mu]}\oplus e_SK_{\mathrm{res},\mu},
  \label{eq:adapter-local-kernel}
\end{equation}
with $e_S=1$ for full-fibre selection. To couple the two fibres with cyclic offset $s\in\mathbb Z_l$, append one ring-valued linking row,
\begin{equation}
  A'_{\mathrm{ad}}
  =
  \begin{pmatrix}
    A'_1&0\\
    0&A'_2\\
    \varepsilon_{i_1}^{\mathsf T}&x^{-s}\varepsilon_{i_2}^{\mathsf T}
  \end{pmatrix}.
  \label{eq:adapter-factor}
\end{equation}
For selected coefficients $\eta_\mu\in R_le_S$, the link imposes
$\eta_1+x^{-s}\eta_2=0$, or $\eta_2=x^s\eta_1$. Residual vectors vanish at the selected coordinates and are therefore unaffected by the link, giving
\begin{equation}
  \ker A'_{\mathrm{ad}}
  =
  R_le_S
  \begin{pmatrix}
    u_{i_1}^{[1]}\\
    x^su_{i_2}^{[2]}
  \end{pmatrix}
  \oplus e_SK_{\mathrm{res},1}
  \oplus e_SK_{\mathrm{res},2}.
  \label{eq:adapter-kernel}
\end{equation}

The coordinate attachments insert the two components at cokernel coordinates $j_1$ and $j_2$. Hence a coefficient $\eta\in R_le_S$ has canonical image
\begin{equation}
  \left[
    \bigl(\eta u_{i_1}^{[1]}\otimes\varepsilon_{j_1},0\bigr)
    \oplus
    \bigl(x^s\eta u_{i_2}^{[2]}\otimes\varepsilon_{j_2},0\bigr)
  \right].
\end{equation}
For full-fibre selection, $\eta=x^a$ gives the $l$ commuting canonical PPMs
\begin{equation}
  M_a^{\mathrm{cyc}}
  =
  \bar q_{i_1,j_1;a}^{[1],\mathrm{cyc}}\,
  \bar q_{i_2,j_2;a+s}^{[2],\mathrm{cyc}},
  \qquad a\in\mathbb Z_l.
  \label{eq:adapter-equal-readout}
\end{equation}
A common CRT selection $S$ instead retains $\sum_{c\in S}\deg g_c$ independent CRT-adapted combinations of these products. Nonzero logical images of $e_SK_{\mathrm{res},1}\oplus e_SK_{\mathrm{res},2}$ are measured in addition.

\paragraph{Unequal lifts.}
Now let the two blocks have odd lift sizes $l_1$ and $l_2$, and set $e=\gcd(l_1,l_2)$ and $m_\mu=l_\mu/e$. Since the two fibres are defined over different rings, we restrict each selected coefficient to an $e$-periodic subspace and couple these common-period degrees of freedom after binary expansion. We restrict this unequal-lift construction to full-fibre addressing.

For block $\mu$, append the period row to the fibre selector,
\begin{equation}
  A'_\mu=
  \begin{pmatrix}
    A^{[\mu]}\\
    H_{L,i_\mu}^{[\mu],\mathrm{fib}}\\
    (x^e+1)\varepsilon_{i_\mu}^{\mathsf T}
  \end{pmatrix},
  \label{eq:adapter-period-checks}
\end{equation}
omitting the last row when $e=l_\mu$. Writing the selected coefficient as
$\eta_\mu=\sum_{a=0}^{l_\mu-1}\widetilde\eta_{\mu,a}x^a$,
the condition $(x^e+1)\eta_\mu=0$ is equivalent to
$\widetilde\eta_{\mu,a+e}=\widetilde\eta_{\mu,a}$ modulo $l_\mu$. Thus the first $e$ coefficients repeat $m_\mu$ times. Define
\begin{equation}
  p_\mu(x)=\sum_{a=0}^{e-1}\widetilde\eta_{\mu,a}x^a\in
  R_e=\mathbb F_2[x]/(x^e+1),
  \qquad
  h_\mu=\sum_{r=0}^{m_\mu-1}x^{re},
  \qquad
  \eta_\mu=h_\mu p_\mu .
  \label{eq:adapter-periodic-coeff}
\end{equation}
Every $p_\mu\in R_e$ is allowed.

After binary expansion, couple the first $e$ coefficients of the repeating blocks. With $P_e(s)_{a,a+s}=1$, take
\begin{equation}
  L_1=\varepsilon_{i_1}^{\mathsf T}\otimes(I_e\mid0),
  \qquad
  L_2=\varepsilon_{i_2}^{\mathsf T}\otimes(P_e(-s)\mid0),
  \label{eq:adapter-link-binary}
\end{equation}
where each zero block has $l_\mu-e$ columns. The link imposes $p_2=x^sp_1$. Writing $p=p_1$, the complete kernel is therefore
\begin{equation}
  \ker A'_{\mathrm{ad}}
  =
  \left\{
    \begin{pmatrix}
      h_1p\,u_{i_1}^{[1]}\\
      h_2x^sp\,u_{i_2}^{[2]}
    \end{pmatrix}
    :p\in R_e
  \right\}
  \oplus K_{\mathrm{res},1}\oplus K_{\mathrm{res},2}.
  \label{eq:adapter-kernel-explicit}
\end{equation}
The residual vectors are unaffected because their selected coordinates vanish. Conversely, every vector in Eq.~\eqref{eq:adapter-kernel-explicit} satisfies the two local equations, the period constraints, and the link equation, so the expression is exact.

For equal lifts, $A'_{\mathrm{ad}}$ remains an $R_l$-linear factor and can be thickened directly using Sec.~\ref{sm:selection-attachment}. For unequal lifts, the common-period link is introduced after binary expansion, so the resulting binary factor is thickened with a binary rank-one tube, again using a normalized coordinate $t_q=1$. The two data attachments are the direct sum of the corresponding single-block coordinate maps, while the linking rows have no data attachment. The selected joint kernel is therefore preserved exactly as in Eq.~\eqref{eq:sm-tube-kernel}. Exchanging $X$ and $Z$ gives the dual construction.

Finally, define the residue-class products
\begin{equation}
  P_a^{[\mu]}
  =
  \prod_{r=0}^{m_\mu-1}
  \bar q_{i_\mu,j_\mu;\,a+re}^{[\mu],\mathrm{cyc}},
  \qquad a\in\mathbb Z_e .
  \label{eq:adapter-residue-products}
\end{equation}
Taking $p=x^a$ in Eq.~\eqref{eq:adapter-kernel-explicit} gives the $e$ commuting canonical PPMs
\begin{equation}
  M_a=P_a^{[1]}P_{a+s}^{[2]},
  \qquad a\in\mathbb Z_e.
  \label{eq:pauli-products}
\end{equation}
For equal lifts, $m_1=m_2=1$ and this reduces to Eq.~\eqref{eq:adapter-equal-readout}. For example, if $l_1=33$ and $l_2=11$, then $e=11$: $P_a^{[1]}$ is a product of three cyclic operators separated by $11$ sites in the first fibre, while $P_{a+s}^{[2]}$ is a single cyclic operator in the second block.

As in the single-block construction, nonzero logical images of the residual kernels are measured in addition to the canonical adapter outputs. The derivation determines the complete measured logical subspace but does not by itself establish preservation of the data-code distance. Unequal-lift adapters additionally require a separate physical-layout analysis of the binary common-period interface.

\subsection{Scope of parallel addressability}
\label{sm:addressability-scope}

The algebraic parallel surgery presented here does not provide arbitrary
logical addressability. The space of measured logical operators must be
invariant under the common cyclic translation. Since
$f_1$ and $\partial_1$ are $R_l$-linear, the measured logical space
$\mathcal M=[f_1\ker\partial_1]$ satisfies $T\mathcal M=\mathcal M$,
where $T$ shifts every logical by one position. Every translate of
a measured operator must therefore belong to the measured subgroup.

\begin{figure}[htbp]
  \centering
  \includegraphics[width=\textwidth]{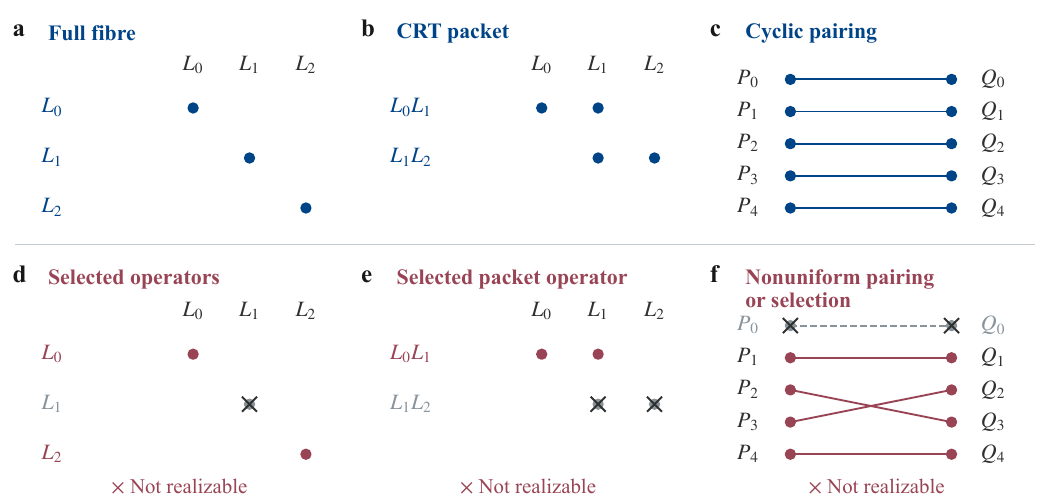}
  \caption{\textbf{Scope of parallel addressability.}
    The operators $L_a$ form one cyclic logical $Z$ basis ($l=3$). Operators $P_a$ and
    $Q_a$ label two such bases ($l=5$). Colored circles mark factors in the
    requested PPMs. Links denote joint products. Crossed gray circles mark
    operators requested to be omitted.
    (a--c) Supported canonical readouts: a full fibre, a complete quadratic
    CRT packet shown through two independent generators, and all pairs
    $P_aQ_a$.
    (d,e) Unsupported selections: omitting $L_1$ or retaining only the packet
    operator $L_0L_1$.
    (f) Nonuniform pairing or selection: the swapped pairs $P_2Q_3$ and
    $P_3Q_2$, or the omitted pair $P_0Q_0$, each independently break
    invariance under the fixed common cyclic translation. The red
    $\times$ labels mark requests that cannot be realized exactly under
    this symmetry. Crosses on gray circles indicate requested exclusions.
  Only canonical readouts are shown.}
  \label{fig:ppm-addressability}
\end{figure}

All PPMs in Fig.~\ref{fig:ppm-addressability} commute, but only the
requests in (a--c) satisfy this condition. In (b), $L_0L_1$ and $L_1L_2$
span the quadratic CRT packet
$\operatorname{span}_{\mathbb F_2}\{1+x,x+x^2\}\simeq
\mathbb F_2[x]/(1+x+x^2)$ in $R_3$. Their product gives the remaining
translate $L_2L_0$.

Panels (d--f) are not realizable by this construction because required
cyclic translates are missing from the requested logical space.
In (d), the missing operator is $T(L_0)=L_1$.
In (e), it is $T(L_0L_1)=L_1L_2$.
In (f), selection omits the required operator $T(P_4Q_4)=P_0Q_0$,
and nonuniform pairing omits $T(P_1Q_1)=P_2Q_2$.

\section{Shuttling depth and timing}

\paragraph{Logarithmic shuttling depth.}
\label{em:memory-routing}
As in the main text, let $n_a$ denote the largest row or column dimension of the unlifted seed matrices $A,B,A',D$, with the selection rows included in $A'$. Assume that the total monomial weight of every row and column is bounded independently of $n_a$ and $l$. The interactions specified by each seed can then be decomposed into a bounded number of matchings. Each matching acts on at most $O(n_a)$ seed coordinates and can be routed using the divide-and-conquer rearrangement of Ref.~\cite{Xu2024ConstantOverhead}: atoms are separated according to whether their destinations lie in the first or second half of the target positions, moved into disjoint regions, and the procedure is applied recursively to both groups in parallel. Each level reduces the unresolved set by a factor of two and requires only a bounded number of collective moves, giving $O(\log n_a)$ motion stages. This construction assumes sufficient workspace and independent control to manipulate the two branches concurrently. Return paths obey the same bound.

For each matching, first route whole fibres to their destination seed coordinates in $O(\log n_a)$ layers, preserving their internal atom order. Then perform the required cyclic shifts within the routed fibres. Each individual shift takes $O(1)$ collective moves, but the required shifts may differ between fibres.

Label the fibres by $i$ and let $s_i\in\{0,\ldots,l-1\}$ be their required shifts. Assume that selecting any subset of fibres and applying a common cyclic shift costs $O(1)$ motion layers, while the other fibres remain fixed. Write $s_i=\sum_{b=0}^{\lceil\log_2 l\rceil-1}\beta_{ib}2^b$, where $\beta_{ib}\in\{0,1\}$. For each bit position $b$, shift all fibres with $\beta_{ib}=1$ by $2^b$ modulo $l$ in parallel. Each fibre then accumulates exactly its required shift $s_i$. There are $\lceil\log_2 l\rceil$ bit positions and each step costs $O(1)$ layers, giving $O(\log l)$ lift-level depth. These shifts follow the protograph routing, so their depths add to $O(\log n_a+\log l)$ per matching. Required returns can reverse these moves and obey the same bound. The bounded number of matchings preserves this bound for a full round.

The LP product structure repeats each seed operation across the other coordinate in parallel, so these copies add no sequential stages.

\label{em:merged-routing}
The same construction applies to the auxiliary LP code through its factors $A'$ and $D$. The coordinate attachments $f_1$ and $f_0$ preserve the relevant seed coordinate and cyclic index. When the coupled data and auxiliary fibres are aligned, they therefore add only a bounded number of collective move stages. Combining the seed-level rearrangements, lift-level cyclic shifts, attachment, and return paths gives
\begin{equation}
  {\delta_{\rm shuttle}^{\rm mem},\;
    \delta_{\rm shuttle}^{\rm merge}
    =O(\log n_a+\log l).
  }
  \label{eq:memory-proof-bound}
\end{equation}
This matches the general reshuffling bound for $l n_a$ atoms. When only a bounded number of distinct monomials occur across each seed matrix, only a bounded number of cyclic shifts are needed. For each matching, fibres requiring the same shift move together under the same control assumption. The lift-level depth is then $O(1)$, leaving $O(\log n_a)$ total depth.

\paragraph{Constant shuttling depth.}
\label{em:constant-depth}
Suppose, in addition, that the nonzero entries of every data and auxiliary seed, including selection checks, lie on a bounded number of diagonals, allowing cyclic wraparound. Connections along each diagonal have the same displacement between seed coordinates. For each monomial on that diagonal, the shifts within fibres are also identical. Under the assumed layout and controls, these interactions can therefore be implemented by a bounded number of collective moves, including those needed at cyclic boundaries. Since both the number of diagonals and the number of distinct monomials are bounded independently of $n_a$ and $l$, protograph routing and the shifts within fibres both take $O(1)$ depth. Thus $\delta_{\rm shuttle}^{\rm mem},\delta_{\rm shuttle}^{\rm merge}=O(1)$, including attachment and return paths.

\paragraph{Physical implementation.}
\label{em:three-blocks}
The simulations use the layouts in the main text. For algebraic surgery, the data and auxiliary codes occupy separate two-dimensional arrays of cyclic fibres. The $l=9$ fibre is placed on a $3\times3$ grid with consecutive indices along each row. For $l=15,33$, the placement $a\mapsto(a\bmod3,a\bmod(l/3))$ gives $3\times5$ and $3\times11$ grids. Syndrome ancillas move between gate locations. The revised sparse auxiliary sweeps also temporarily park inactive auxiliary data off the gate line, as described below. Each crossed AOD pair consists of two perpendicular AODs controlling a rectangular trap grid. We assign one or two pairs to each code, while memory uses only the pairs assigned to the data code.

Each syndrome ancilla performs all interactions belonging to its merged check before measurement. In particular, an added $Z$-check ancilla interacts through both $\partial_1^\dagger$ and $f_1^\dagger$, while an original $X$-check ancilla interacts through both $H_X$ and $f_0$. There is no measurement or reset between these interactions. Operations on the data and auxiliary arrays may overlap when they use independent controls and respect the required ordering of gates on shared atoms.

For the graph-based comparison, the auxiliary atoms form a one-dimensional register beside the unchanged LP data array. Check atoms are rearranged between successive matchings, retaining their current order until the next rearrangement. Attachment also requires original $X$-check ancillas to visit auxiliary qubits and added $Z$-check ancillas to visit data qubits. These motions include rearrangement outside the data array and visits grouped by physical column. With two auxiliary AOD pairs, the two groups in the divide-and-conquer rearrangement can move concurrently. Both internal auxiliary interactions and attachment are included in the reported times.

\paragraph{Timing and comparison.}
\label{em:timing}
We use the shuttling model of the main text, with $12\,\mu\mathrm m$ site spacing, $2\,\mu\mathrm m$ separation during entangling gates, and $1\,\mu\mathrm s$ per entangling pulse. A move of length $L$ takes $\tau(L)=2\sqrt{L/a_{\max}}$, with $a_{\max}=5500\,\mathrm{m\,s^{-2}}$. Simultaneous moves are timed by the largest displacement. The round time is then calculated from the duration-weighted critical path, including the required gate ordering and scheduled returns to the home fibres.

Equal shuttling depth therefore need not imply equal duration: at comparable atom spacing, a compact two-dimensional layout can shorten travel distances relative to a one-dimensional register and thereby reduce shuttling time.

Table~\ref{tab:all-times} combines separate $X$ and $Z$ extraction, a nonparallel $X+Z$ round, and the average time for a full round of parallel $X/Z$ extraction. The first three timing columns for each allocation use AOD pairs shared between the two bases within each code, with $X$ followed by $Z$ in each bank. The parallel column uses the independent basis banks described below. The data and auxiliary arrays can proceed independently when their controls and gate ordering permit. They need not both finish one type of check before either starts the other.
The values include shuttling, docking, entangling gates, and returns. In the algebraic layout, entire blocks move along transverse passing lanes and detours around parked check blocks. The $f_0$ and $f_1$ interface checks travel in a $6\,\mu\mathrm m$ side lane, stay $2\,\mu\mathrm m$ from their targets on the side opposite ordinary checks, and reverse this route on return. The times exclude readout, reset, finite trap transfer time, and collision avoidance within blocks for the LP layout. The separate $X$, $Z$, and nonparallel $X+Z$ entries describe one extraction round before repetition for fault tolerance. The parallel column gives the average over $r$ rounds.

The algebraic construction addresses all $l$ canonical logical $Z$ operators in one fibre, whereas each timed graph gadget targets one fixed canonical logical $Z$ operator. The table reports these respective measurements under the same transport model and stated AOD allocation. The three lifts use different seed and auxiliary-factor choices. Their timings describe the finite instances considered here, while the depth bounds follow from the structural assumptions above.

To reduce the cost of zero entries in $A'$, the revised auxiliary sweep carries check columns through successive cyclic matchings and temporarily parks inactive auxiliary data columns $6\,\mu\mathrm m$ off the gate line. Each data column returns before its next nonzero interaction. The compiler selects this route separately for the $X$ and $Z$ sweeps only when it is faster than the original route. All parking and return motions and their AOD occupancy are included, without adding AOD pairs to the nonparallel schedules in Table~\ref{tab:all-times}. Checks return to their home fibres but can retain a permutation within each fibre. The memory baseline uses the same endpoint rule, local cyclic-shift routes, and whole-block return detours. It does not restore sheet labels after each sector. Fresh check assignments can change only after readout and reset. The parallel schedules restore labelled positions after the final round. Memory has dense seed factors, so it requires no sparse auxiliary-data parking. The graph attachment routes still require labelled LP returns, which remain included in their timings. At $l=33$, the separate algebraic $Z$ times fall from $35.38$ to $29.36\,\mathrm{ms}$ with one pair per code and from $23.56$ to $18.85\,\mathrm{ms}$ with two. The corresponding $X$ times are $25.48$ and $16.34\,\mathrm{ms}$. These schedules retain the stated exclusions. Collision avoidance for simultaneous routes across banks is not fully certified.

\begin{table*}[t]
  \centering
  \small
  \setlength{\tabcolsep}{6pt}
  \caption{Shuttling and gate times (ms). $X$, $Z$, and $X+Z$ (no parallel) use $p$ crossed AOD pairs per code. Parallel $X+Z$ uses $2p$ per code in independent $X/Z$ banks, totaling $2p$ for memory and $4p$ for surgery. Parallel entries show the average time for a full $X+Z$ round, $T_{XZ}^{(r)}/r$, including startup and final restoration. We use $r=8,10,20$ at $l=9,15,33$, with $20$ taken from the distance upper bound. Nonparallel $X+Z$ is jointly compiled and need not equal the sum of the separate $X$ and $Z$ times. Readout, reset, and trap transfer durations are excluded. Dashes denote unevaluated schedules. Algebraic surgery measures a full canonical fibre, while each graph gadget measures one observable.}
  \label{tab:all-times}
  \begin{tabular}{cl rrrr@{\qquad}rrrr}
    \toprule
    &&\multicolumn{4}{c}{$p=1$}&\multicolumn{4}{c}{$p=2$}\\
    \cmidrule(lr){3-6}\cmidrule(lr){7-10}
    $l$&Protocol&$X$&$Z$&\shortstack{$X+Z$\\no parallel}&\shortstack{$X+Z$\\parallel}&$X$&$Z$&\shortstack{$X+Z$\\no parallel}&\shortstack{$X+Z$\\parallel}\\
    \midrule
    9&Memory&19.16&19.16&38.32&22.113&11.91&11.91&23.82&13.761\\
    9&Algebraic fibre&20.42&19.16&39.58&22.374&13.17&11.91&25.09&14.270\\
    9&Graph-WY&30.16&19.67&47.21&\textemdash&21.36&13.98&35.34&\textemdash\\
    9&Graph-IGND&29.89&19.67&46.64&\textemdash&21.04&13.71&34.75&\textemdash\\
    \addlinespace
    15&Memory&17.95&18.14&36.08&21.253&11.67&11.77&23.44&13.501\\
    15&Algebraic fibre&19.21&18.14&37.35&21.409&12.94&11.77&24.70&13.808\\
    15&Graph-WY&43.03&27.00&70.04&\textemdash&33.36&21.58&54.93&\textemdash\\
    15&Graph-IGND&33.49&26.95&60.44&\textemdash&26.56&21.51&48.06&\textemdash\\
    \addlinespace
    33&Memory&23.14&23.73&46.87&25.947&15.02&15.32&30.35&16.642\\
    33&Algebraic fibre&25.48&29.36&54.84&30.430&16.34&18.85&34.60&19.728\\
    33&Graph-WY&67.37&42.98&110.34&\textemdash&50.07&32.69&82.76&\textemdash\\
    33&Graph-IGND&55.97&42.92&98.89&\textemdash&42.02&32.71&74.72&\textemdash\\
    \bottomrule
  \end{tabular}
\end{table*}

\paragraph{Parallel $X/Z$ extraction over $r$ rounds.}
\label{supp:parallel-xz-timing}
We also compile the staggered schedule of Ref.~\cite{StrikisBrowneBeverland2026}, as implemented in the animation. Sweeps of $X$ and $Z$ checks overlap on complementary data regions. Successive rounds are staggered, preserving the required order on shared data. Each code now has independent $X$ and $Z$ AOD banks. An allocation of $p=1$ or $2$ crossed pairs per basis per code therefore uses $2p$ pairs for memory and $4p$ pairs for algebraic fibre surgery, twice the AOD allocation used for the corresponding nonparallel columns in Table~\ref{tab:all-times}. The check and data qubit counts are unchanged. The graph schedules in that table are not recompiled with this parallel construction.

Permutations within each fibre are retained between sweeps and rounds. Assignments of checks to atoms change only after readout and reset, and remain fixed during each check measurement. The sparse $A'$ sweeps use the same timed parking motions as above, reserving each parked data column for its full excursion. The compiler preserves dependencies on atoms, AOD pairs, and shared data regions. It restores the original labelled atom positions only after the final readouts. We set the durations of readout and reset to zero, consistently with Table~\ref{tab:all-times}. These operations impose ordering but contribute no elapsed time. Startup, the final completion of both bases, and labelled restoration are all included in the reported total.

For $r$ rounds, let $T_{XZ}^{(r)}$ denote this full elapsed time. The parallel column reports the average time per complete $X/Z$ round,
\begin{equation}
  t_{XZ}^{(r)}=\frac{T_{XZ}^{(r)}}{r}.
  \label{eq:parallel-xz-normalization}
\end{equation}
Each round includes extraction of both check bases. We use $r=d=8$ and $10$ for $l=9$ and $15$, respectively. For $l=33$, we use $r=20$, matching the distance bound and the circuit benchmark. The stated $d\leq20$ does not establish equality. The parallel column of Table~\ref{tab:all-times} uses the total elapsed time for the specified number of rounds, including startup and final restoration. For algebraic fibre surgery, the average times for a full round are $22.374$, $21.409$, and $30.430\,\mathrm{ms}$ with one pair per basis per code, and $14.270$, $13.808$, and $19.728\,\mathrm{ms}$ with two, for $l=9,15,33$. These estimates describe transport and gates under the stated layout assumptions. They do not establish circuit distance or logical error rates for the new extraction schedule.

\section{Three code examples}
\subsection{$[[306,52,8]]$ LP data code with $l=9$}
\label{app:lp9-example}
We give an explicit example of the cyclic orbit and CRT power
logical bases, fibre surgery, and an adapter between two blocks. A monomial tube
preserves the distance of a $[[306,52,8]]$ data block while using
279 auxiliary qubits. The resulting merged code has parameters
$[[585,41,8]]$. The auxiliary-code cost is minimal within the construction specified below, which fixes the factor and uses attachment through a single coordinate.

\subsubsection{Data code and logical bases}
Work over $R_9=\mathbb F_2[x]/(x^9+1)$ and take
\begin{equation}
  A=
  \begin{pmatrix}
    1&1&1&1&1\\
    1&x^5&x^2&x^6&x\\
    1&x^8&x^4&x^5&x^2
  \end{pmatrix},\qquad B=A^\dagger.
  \label{eq:s-lp9-seed}
\end{equation}
The adjoint transposes the matrix and sends $x$ to $x^{-1}$. With the
qubit ordering $(A_1\otimes B_0)\oplus(A_0\otimes B_1)$, the checks are
\begin{equation}
  H_X=(A\otimes I_5\mid I_3\otimes A^\dagger),\qquad
  H_Z=(I_5\otimes A\mid A^\dagger\otimes I_3).
  \label{eq:s-lp9-data-checks}
\end{equation}
All tensor products in this section are over $R_9$. We number coordinates from zero and use the lift convention $P(s)_{a,a+s}=1$, with sheet indices
modulo nine. Thus $x^a$ times a polynomial column is represented by
column $-a$ of its lift. The block has $n=9(5^2+3^2)=306$ physical
qubits. Both binary check ranks are 127, giving $k=52$. The distances
in both Pauli sectors are eight.

The factorization and corresponding CRT idempotents are
\begin{align}
  x^9+1&=(1+x)(1+x+x^2)(1+x^3+x^6),
  \label{eq:s-lp9-crt-factors}\\
  \epsilon_0&=\sum_{a=0}^8x^a,\qquad
  \epsilon_1=x+x^2+x^4+x^5+x^7+x^8,\qquad
  \epsilon_2=x^3+x^6.
  \label{eq:s-lp9-idempotents}
\end{align}
The component fields $E_c$ have degrees $(d_0,d_1,d_2)=(1,2,6)$,
and the component ranks of $A$ are $(1,3,3)$. Elimination from left to right
gives the common information set $J=J^\dagger=\{3,4\}$, hence a
$2\times2$ array of canonical fibres. To make the gluing explicit, label
the three factors in Eq.~\eqref{eq:s-lp9-crt-factors} by $g_0,g_1,g_2$
in the displayed order. Put $E_c=\mathbb F_2[x]/(g_c)$ and
$A_c=A\bmod g_c$. The sets of columns without pivots are
$J_0=\{1,2,3,4\}$ and $J_1=J_2=\{3,4\}$. For each $i\in J_c$,
the systematic packet generator is the column vector
\begin{equation}
  u_i^{(c)}\in\ker_{E_c}A_c\subseteq E_c^5,\qquad
  (u_i^{(c)})_j=
  \delta_{i,j}\quad(j\in J_c),\qquad (u_i^{(c)})_i=1.
  \label{eq:s-lp9-packet-columns}
\end{equation}
Here $i$ labels the free coordinate that is set to one. All other free
coordinates are set to zero, and the pivot coordinates are determined
by $A_cu_i^{(c)}=0$. In particular, the two generators with labels in the
common set $J=\{3,4\}$ are
\begin{align}
  u_3^{(0)}&=(1,0,0,1,0)^{\mathsf T},\nonumber\\
  u_4^{(0)}&=(1,0,0,0,1)^{\mathsf T},\nonumber\\
  u_3^{(1)}&=(1,x,x,1,0)^{\mathsf T},\nonumber\\
  u_4^{(1)}&=(1+x,1+x,1,0,1)^{\mathsf T},\nonumber\\
  u_3^{(2)}&=(1+x^3+x^5,x^2+x^5,x^2+x^3,1,0)^{\mathsf T},\nonumber\\
  u_4^{(2)}&=(1+x^2+x^3+x^4+x^5,x^3+x^5,x^2+x^4,0,1)^{\mathsf T}.
  \label{eq:s-lp9-packet-kernels}
\end{align}
Each entry is the polynomial representative of degree less than
$\deg g_c$, with arithmetic in $E_c=\mathbb F_2[x]/(g_c)$.
For $c=1,2$, the pivot columns are $0,1,2$. The displayed first three
coordinates solve the reduced equations for each choice of free
coordinate. In packet zero, the sole independent equation is
$z_0+z_1+z_2+z_3+z_4=0$. Its other two systematic generators are
$u_1^{(0)}=(1,1,0,0,0)^{\mathsf T}$ and
$u_2^{(0)}=(1,0,1,0,0)^{\mathsf T}$. The labels $1,2$ are absent from
$J_1,J_2$ and therefore do not define full canonical fibres.

The packet vectors $u_i^{(c)}$ belong to different fields. To form a global
vector, let $\widetilde u_i^{(c)}\in R_9^5$ denote the entrywise polynomial
representative of $u_i^{(c)}$ displayed above. The CRT reconstruction is
\begin{equation}
  u_i=\sum_{c=0}^2\epsilon_c\widetilde u_i^{(c)}\in\ker_{R_9}A,
  \qquad u_i\bmod g_c=u_i^{(c)},\qquad i\in\{3,4\}.
  \label{eq:s-lp9-packet-gluing}
\end{equation}
All products in this sum are reduced modulo $x^9+1$.
It is independent of the polynomial representatives because
$\epsilon_cg_c=0$ in $R_9$. Since $\sum_c\epsilon_c=1$, the
normalization becomes $(u_i)_j=\delta_{i,j}$ for $i,j\in\{3,4\}$.
For example,
the first coordinates of the two columns reduce as
\begin{align}
  (u_3)_0&=\epsilon_0+\epsilon_1+(1+x^3+x^5)\epsilon_2
  =x^2+x^6+x^8,\nonumber\\
  (u_4)_0&=\epsilon_0+(1+x)\epsilon_1
  +(1+x^2+x^3+x^4+x^5)\epsilon_2
  =1+x+x^3+x^7+x^8.
  \label{eq:s-lp9-gluing-entries}
\end{align}
Applying the same entrywise sum to all coordinates gives the two
global generators
\begin{equation}
  u_3=
  \begin{pmatrix}
    x^2+x^6+x^8\\1+x^3+x^6+x^8\\x^2+x^3\\1\\0
  \end{pmatrix},\qquad
  u_4=
  \begin{pmatrix}
    1+x+x^3+x^7+x^8\\x+x^2+x^3+x^4+x^7+x^8\\x^2+x^4\\0\\1
  \end{pmatrix}.
  \label{eq:s-lp9-generators}
\end{equation}
These satisfy $Au_i=0$ and
$\varepsilon_j^{\mathsf T}u_i=\delta_{i,j}$ for $i,j\in J$, where
$\varepsilon_j$ denotes a standard coordinate column. Coordinate
representatives $v_j=\varepsilon_j$ then give canonical $Z$ classes
$q_{i,j}=[(u_i\otimes\varepsilon_j,0)]$. Their cyclic orbit basis is
$\bar q_{i,j;a}^{\mathrm{cyc}}=x^a q_{i,j}$, $a=0,\ldots,8$, with dual $X$
representatives $x^a(\varepsilon_i\otimes u_j,0)$.
Choosing the power basis $1,x,\ldots,x^{d_c-1}$ in each field
$E_c$ defines the CRT power basis
\begin{equation}
  \bar q_{i,j;b}^{\mathrm{CRT}(c)}=\epsilon_c x^bq_{i,j},
  \qquad 0\le b<d_c.
  \label{eq:s-lp9-bases}
\end{equation}
For each fibre the latter consists of one, two, and six binary operators
in the three packets. To distinguish the two index ranges, use $a$ for
the cyclic coordinate, $0\le a<l$, and $b$ for the coordinate
within packet $c$, $0\le b<d_c$. For fixed $(i,j)$, collect the cyclic
classes into $\mathbf q_{i,j}^{\mathrm{cyc}}=(\bar q_{i,j;a}^{\mathrm{cyc}})_{a=0}^{l-1}$
and the CRT classes into $\mathbf q_{i,j}^{\mathrm{CRT}}$, ordered first
by $c$ and then by $b$. Both are columns of logical classes. Define
$r(c,b)=\sum_{h<c}d_h+b$. The basis changes are
\begin{align}
  \mathbf q_{i,j}^{\mathrm{CRT}}&=T_l\mathbf q_{i,j}^{\mathrm{cyc}},&
  (T_l)_{r(c,b),a}&=[x^a](\epsilon_cx^b\bmod(x^l+1)),\nonumber\\
  \mathbf q_{i,j}^{\mathrm{cyc}}&=T_l^{-1}\mathbf q_{i,j}^{\mathrm{CRT}},&
  (T_l^{-1})_{a,r(c,b)}&=[x^b](x^a\bmod g_c).
  \label{eq:s-basis-change-rule}
\end{align}
Here $[x^a]$ denotes coefficient extraction and all matrix arithmetic
is over $\mathbb F_2$. The inverse formula follows by expanding
$x^a=\sum_c\epsilon_c(x^a\bmod g_c)$ in the CRT power basis.
These equations transform the basis elements themselves. Coefficient
columns of a fixed logical class instead obey
$\lambda^{\mathrm{CRT}}=T_l^{-\mathsf T}\lambda^{\mathrm{cyc}}$.
At lift nine the row order is
$(0,0),(1,0),(1,1),(2,0),\ldots,(2,5)$, giving
\begingroup
\setlength{\arraycolsep}{4pt}
\begin{equation}
  T_{9}=\left(
    \begin{array}{ccccccccc}
      1 & 1 & 1 & 1 & 1 & 1 & 1 & 1 & 1 \\
      \hline
      0 & 1 & 1 & 0 & 1 & 1 & 0 & 1 & 1 \\
      1 & 0 & 1 & 1 & 0 & 1 & 1 & 0 & 1 \\
      \hline
      0 & 0 & 0 & 1 & 0 & 0 & 1 & 0 & 0 \\
      0 & 0 & 0 & 0 & 1 & 0 & 0 & 1 & 0 \\
      0 & 0 & 0 & 0 & 0 & 1 & 0 & 0 & 1 \\
      1 & 0 & 0 & 0 & 0 & 0 & 1 & 0 & 0 \\
      0 & 1 & 0 & 0 & 0 & 0 & 0 & 1 & 0 \\
      0 & 0 & 1 & 0 & 0 & 0 & 0 & 0 & 1
  \end{array}\right),\qquad
  T_{9}^{-1}=\left(
    \begin{array}{c|cc|cccccc}
      1 & 1 & 0 & 1 & 0 & 0 & 0 & 0 & 0 \\
      1 & 0 & 1 & 0 & 1 & 0 & 0 & 0 & 0 \\
      1 & 1 & 1 & 0 & 0 & 1 & 0 & 0 & 0 \\
      1 & 1 & 0 & 0 & 0 & 0 & 1 & 0 & 0 \\
      1 & 0 & 1 & 0 & 0 & 0 & 0 & 1 & 0 \\
      1 & 1 & 1 & 0 & 0 & 0 & 0 & 0 & 1 \\
      1 & 1 & 0 & 1 & 0 & 0 & 1 & 0 & 0 \\
      1 & 0 & 1 & 0 & 1 & 0 & 0 & 1 & 0 \\
      1 & 1 & 1 & 0 & 0 & 1 & 0 & 0 & 1
  \end{array}\right).
  \label{eq:s-lp9-basis-change-matrices}
\end{equation}
\endgroup
Horizontal rules in $T_9$ separate CRT packets. Vertical rules in
$T_9^{-1}$ separate the corresponding packet coordinates. For example,
\begin{align}
  \bar q_{i,j;0}^{\mathrm{CRT}(0)}=\sum_{a=0}^8\bar q_{i,j;a}^{\mathrm{cyc}},\ \
  \bar q_{i,j;0}^{\mathrm{CRT}(2)}&=\bar q_{i,j;3}^{\mathrm{cyc}}+\bar q_{i,j;6}^{\mathrm{cyc}},\ \ \bar q_{i,j;0}^{\mathrm{cyc}}=\bar q_{i,j;0}^{\mathrm{CRT}(0)}
  +\bar q_{i,j;0}^{\mathrm{CRT}(1)}
  +\bar q_{i,j;0}^{\mathrm{CRT}(2)}.&&
  \label{eq:s-lp9-basis-change-examples}
\end{align}
Addition of these $Z$-type classes corresponds to multiplication of
their commuting logical Pauli representatives. The dual $X$ basis
transforms by $T_9^{-\mathsf T}$ to preserve Pauli pairing. There are
36 canonical pairs and 16 additional residual pairs, all of the latter
in $E_0$.

We write $\bar Z_{i,j;a}^{\mathrm{cyc}}$ for a chosen $Z$-type Pauli
representative of the logical class $\bar q_{i,j;a}^{\mathrm{cyc}}$.

An explicit selective measurement is the packet of degree six $S=\{2\}$.
A basis of its canonical readout space consists of the products
\begin{equation}
  \begin{gathered}
    \bar Z_{i,j;3}^{\mathrm{cyc}}\bar Z_{i,j;6}^{\mathrm{cyc}},\quad
    \bar Z_{i,j;4}^{\mathrm{cyc}}\bar Z_{i,j;7}^{\mathrm{cyc}},\quad
    \bar Z_{i,j;5}^{\mathrm{cyc}}\bar Z_{i,j;8}^{\mathrm{cyc}},\\
    \bar Z_{i,j;6}^{\mathrm{cyc}}\bar Z_{i,j;0}^{\mathrm{cyc}},\quad
    \bar Z_{i,j;7}^{\mathrm{cyc}}\bar Z_{i,j;1}^{\mathrm{cyc}},\quad
    \bar Z_{i,j;8}^{\mathrm{cyc}}\bar Z_{i,j;2}^{\mathrm{cyc}}.
  \end{gathered}
  \label{eq:s-lp9-packet-parities}
\end{equation}
Their outcomes are recovered as parities of the outcomes of added checks.
The carving does not prescribe an ordered basis of raw readouts.
These six products are independent and do not determine any individual cyclic
$\bar Z_{i,j;a}^{\mathrm{cyc}}$. An ideal measurement of these six products leaves
three logical degrees of freedom within the canonical fibre of nine qubits. Moreover, this packet has no residual readout because
$\epsilon_2K_{\mathrm{res}}=0$.

\subsubsection{Algebraic surgery construction and thickening}
To measure the full fibre $(i,j)=(3,3)$, choose
\begin{equation}
  H_L=\varepsilon_4^{\mathsf T}=(0,0,0,0,1),\qquad
  A'=
  \begin{pmatrix}A\\H_L
  \end{pmatrix}.
  \label{eq:s-lp9-carving}
\end{equation}
The residual factor kernel is
$K_{\mathrm{res}}=\operatorname{span}_{\mathbb F_2}
\{\epsilon_0(\varepsilon_0+\varepsilon_1),
\epsilon_0(\varepsilon_0+\varepsilon_2)\}$, so
$\ker A'=R_9u_3\oplus K_{\mathrm{res}}$ has binary dimension 11.
The coordinate check therefore retains nine canonical and two residual
readouts, all of which are included in the following construction.

Choose the $3\times4$ tube with three monomials in every row,
\begin{equation}
  D=
  \begin{pmatrix}
    1&1&1&0\\
    0&x^2&x^6&1\\
    x^8&0&x^5&1
  \end{pmatrix},\qquad
  t=\Delta^{-1}
  \begin{pmatrix}
    x^2+x^5+x^6\\x^5+x^6+x^8\\x^2+x^8\\\Delta
  \end{pmatrix},\qquad
  \Delta=x+x^5+x^7.
  \label{eq:s-lp9-tube}
\end{equation}
The minor obtained by deleting column three has determinant $\Delta$,
a unit with $\Delta^{-1}=x+x^2+x^3+x^5+x^6$ in $R_9$.
Direct multiplication gives $Dt=0$ and $t_3=1$, so
$\ker D=R_9t$ and every CRT component has kernel dimension one.
We therefore couple at $q=3$.

The auxiliary boundaries of $\mathrm{LP}(A',D)$ are
\begin{equation}
  \partial_1=
  \begin{pmatrix}A'\otimes I_4\\I_5\otimes D
  \end{pmatrix},\qquad
  \partial_0=(I_4\otimes D\mid A'\otimes I_3),
  \label{eq:s-lp9-boundaries}
\end{equation}
of sizes $31\times20$ and $12\times31$, respectively.
Writing $e_3\in R_9^4$ for the fourth tube coordinate and
$W=\varepsilon_3 e_3^{\mathsf T}$, the attachment maps are
\begin{equation}
  f_1=
  \begin{pmatrix}I_5\otimes W\\0_{9\times20}
  \end{pmatrix},\qquad
  f_0=\bigl((I_3\mid0_{3\times1})\otimes W\mid0_{15\times15}\bigr).
  \label{eq:s-lp9-wiring}
\end{equation}
They satisfy $H_Xf_1=f_0\partial_1$ and
$\partial_0\partial_1=0$. Hence the merged checks commute. Moreover,
$\ker\partial_1=(R_9u_3\oplus K_{\mathrm{res}})\otimes R_9t$,
and evaluation at $q=3$ recovers the original carved kernel. Products
of the added $Z$ checks indexed by this kernel cancel all auxiliary
support and span the readout space generated by the nine target
canonical operators and two residual operators. Thus the tube preserves the complete readout
space, leaving $52-11=41$ logical pairs.

\subsubsection{Adapter between two blocks}
For two copies of the same data code, couple fibres $(3,3)$ and $(4,4)$
with offset $\alpha=1$. Define
\begin{equation}
  A'_1=
  \begin{pmatrix}A\\\varepsilon_4^{\mathsf T}
  \end{pmatrix},\qquad
  A'_2=
  \begin{pmatrix}A\\\varepsilon_3^{\mathsf T}
  \end{pmatrix},\qquad
  A'_{\mathrm{ad}}=
  \begin{pmatrix}A'_1&0\\0&A'_2\\
    \varepsilon_3^{\mathsf T}&x\varepsilon_4^{\mathsf T}
  \end{pmatrix}.
  \label{eq:s-lp9-adapter-factor}
\end{equation}
The link imposes $\eta_1+x\eta_2=0$, giving
\begin{equation}
  \ker A'_{\mathrm{ad}}=
  R_9
  \begin{pmatrix}u_3^{[1]}\\x^{-1}u_4^{[2]}
  \end{pmatrix}
  \oplus(K_{\mathrm{res},1}\oplus K_{\mathrm{res},2}).
  \label{eq:s-lp9-adapter-kernel}
\end{equation}
Thickening with the same $D$ and coupling at $q=3$ measures the nine
products $\bar Z_{33;a}^{[1],\mathrm{cyc}}\bar Z_{44;a-1}^{[2],\mathrm{cyc}}$, together with
four local residual operators. No individual canonical operator is
measured. The $9\times10$ factor gives
$9(9\cdot4+10\cdot3)=594$ auxiliary qubits, 1206 qubits in total,
and $2\cdot52-13=91$ surviving logical pairs. Commutation and the
complete readout space have been verified with the updated tube.
The merged check ranks are 501 and 614.

\subsubsection{Logical error rates}
\label{app:lp9-ler}
\begin{figure}[htbp]
  \centering
  \includegraphics[width=0.9\linewidth]{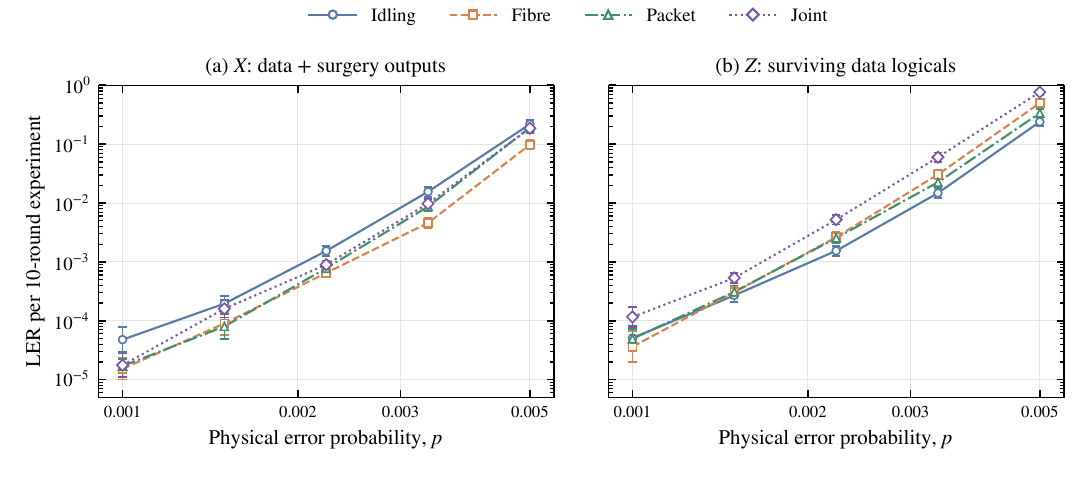}
  \caption{\textbf{Logical error rates under circuit noise for $l=9$.}
    Idling is compared with fibre, packet, and joint surgery.
    (a) Failure of any original data logical or canonical surgery output
    in the $X$ experiment. (b) Failure of any surviving data logical in
    the $Z$ experiment. Each point reports failures divided by shots for
    ten noisy syndrome rounds. Error bars are nominal 95\% Wilson
    intervals and are not adjusted for adaptive stopping.
  Connecting lines guide the eye.}
  \label{fig:lp9-ler}
\end{figure}

Figure~\ref{fig:lp9-ler} compares idling, fibre surgery, packet surgery,
and joint surgery between two copies of the $l=9$ data code.
Each experiment uses ten noisy syndrome rounds, with separate $X$ and $Z$
preparation and readout. The surgery operators are of $X$ type.
The packet experiment uses the multiplier $1+x^3$ to select the CRT union
$S=\{0,1\}$, giving three canonical readouts and two residual readouts.
The joint experiment couples fibre $(3,3)$ in both blocks with zero offset.
Fibre and joint surgery each give nine canonical readouts, together with
two and four residual readouts, respectively.

The simulations use depolarizing circuit noise at five logarithmically
spaced probabilities from $p=0.001$ to $0.005$. The shared setup and
decoder parameters are given in
\hyperref[app:simulation-methods]{Simulation methods and decoder settings}.
The completed dataset contains 9\,436\,691 shots across 40 combinations of
protocol, basis, and error probability, with 3\,178 logical failures.

For the $X$ experiment, a shot fails if any original data logical or any
additional canonical surgery output is wrong. Errors in both categories
are counted once. In the order idling, fibre, packet, and joint surgery,
the numbers of recorded observables are 52, 61, 55, and 113.
For the $Z$ experiment, failure means an error in any surviving data
logical, with 52, 41, 47, and 91 observables, respectively.
Joint surgery uses two data blocks, while the other protocols use one.
Rates are reported per complete experiment of ten rounds.

\subsection{$[[510,76,10]]$ LP data code with $l=15$}
\label{app:lp15-example}
We next construct a $[[510,76,10]]$ data code over
$R_{15}=\mathbb F_2[x]/(x^{15}+1)$. The canonical register again has
four fibres, now containing 15 binary logical pairs each. The remaining
16 pairs are residual. A tube with nine monomial entries produces a
$[[975,59,10]]$ merged code using 465 auxiliary qubits. We use the same qubit ordering, adjoint, and
coordinate conventions as in the example with lift nine, with all sheet
indices reduced modulo 15.

\subsubsection{Data code and logical bases}
The two sizes considered were $2\times4$ and $3\times5$, corresponding
to 300 and 510 physical qubits. The smaller seed with exponent rows
$(0,0,0,0)$ and $(0,1,3,7)$ gives $[[300,66,6]]$. For the worked
construction we choose the seed with larger distance
\begin{equation}
  A=
  \begin{pmatrix}
    1&1&1&1&1\\
    1&x^8&x^{13}&x^7&x^6\\
    1&x^{14}&x^6&x^4&x^{13}
  \end{pmatrix},\qquad B=A^\dagger.
  \label{eq:s-lp15-seed}
\end{equation}
The checks have the form of Eq.~\eqref{eq:s-lp9-data-checks} over
$R_{15}$. Both binary check ranks are 217, giving
$n=15(25+9)=510$ and $k=510-2\cdot217=76$.
The seed search examined all $15^3=3375$ arrays in the normalized
$2\times4$ family and 386 $3\times5$ candidates. Candidates were
screened for the required component ranks and common information set,
connected lifts, and absence of cycles of length four. These searches selected a
small example. They do not establish optimality over all LP seeds.

For the CRT decomposition, the five irreducible factors, in the order used below, are
\begin{align}
  g_0&=1+x,& g_1&=1+x+x^2,& g_2&=1+x+x^4,\nonumber\\
  g_3&=1+x^3+x^4,& g_4&=1+x+x^2+x^3+x^4,&
  x^{15}+1&=\prod_{c=0}^4g_c.
  \label{eq:s-lp15-factors}
\end{align}
Their degrees are $(1,2,4,4,4)$, and the component ranks of $A$ are
$(1,3,3,3,3)$. Elimination from left to right gives free columns
$J_0=\{1,2,3,4\}$ and $J_c=\{3,4\}$ for $c>0$.
The coordinate cokernel representatives are independent at the same
indices, so $J=J^\dagger=\{3,4\}$.

For compact explicit expressions, write
$p_{a_1,\ldots,a_s}=x^{a_1}+\cdots+x^{a_s}$ in $R_{15}$.
The CRT idempotents satisfying $\epsilon_c\bmod g_d=\delta_{cd}$ are
\begin{align}
  \epsilon_0&=\sum_{a=0}^{14}x^a,&
  \epsilon_1&=p_{1,2,4,5,7,8,10,11,13,14},\nonumber\\
  \epsilon_2&=p_{1,2,3,4,6,8,9,12},&
  \epsilon_3&=p_{3,6,7,9,11,12,13,14},\nonumber\\
  \epsilon_4&=p_{1,2,3,4,6,7,8,9,11,12,13,14}.&&
  \label{eq:s-lp15-idempotents}
\end{align}
Let $E_c=\mathbb F_2[x]/(g_c)$ and $A_c=A\bmod g_c$. With the
sets of free columns $J_c$ specified above, the packet generators satisfy
\begin{equation}
  u_i^{(c)}\in\ker_{E_c}A_c\subseteq E_c^5,\qquad
  (u_i^{(c)})_j=\delta_{i,j}\quad(j\in J_c),\qquad (u_i^{(c)})_i=1.
  \label{eq:s-lp15-packet-normalization}
\end{equation}
For the common labels $i=3,4$, the corresponding column vectors are
\begin{align}
  u_3^{(0)}&=(1,0,0,1,0)^{\mathsf T},\nonumber\\
  u_4^{(0)}&=(1,0,0,0,1)^{\mathsf T},\nonumber\\
  u_3^{(1)}&=(1+x,x,0,1,0)^{\mathsf T},\nonumber\\
  u_4^{(1)}&=(x,x,1,0,1)^{\mathsf T},\nonumber\\
  u_3^{(2)}&=(1+x^3,0,x^3,1,0)^{\mathsf T},\nonumber\\
  u_4^{(2)}&=(x^2+x^3,1+x^2,x^3,0,1)^{\mathsf T},\nonumber\\
  u_3^{(3)}&=(x+x^2,1+x+x^2,0,1,0)^{\mathsf T},\nonumber\\
  u_4^{(3)}&=(1+x+x^2,1+x+x^2+x^3,1+x^3,0,1)^{\mathsf T},\nonumber\\
  u_3^{(4)}&=(1+x^2,1+x+x^2,1+x,1,0)^{\mathsf T},\nonumber\\
  u_4^{(4)}&=(x^3,1,x^3,0,1)^{\mathsf T}.
  \label{eq:s-lp15-packet-kernels}
\end{align}
Each displayed entry is a polynomial representative of degree less
than $\deg g_c$ in the corresponding field $E_c$.
For $c=1,\ldots,4$, columns $0,1,2$ are pivots, so the first three
coordinates are fixed by $A_cu_i^{(c)}=0$ once coordinates $3,4$ are
set to $(1,0)$ or $(0,1)$.
Packet zero has the same equation with all coefficients equal to one as at lift nine: the two
displayed columns set coordinates one and two to zero, while
$u_1^{(0)}=(1,1,0,0,0)^{\mathsf T}$ and
$u_2^{(0)}=(1,0,1,0,0)^{\mathsf T}$ are its two additional normalized
kernel generators. Their labels are absent from the other packets.
They account for the residual factor kernel below.

Let $\widetilde u_i^{(c)}\in R_{15}^5$ denote the polynomial representative
of the packet vector $u_i^{(c)}$ just displayed. The global generators are
\begin{equation}
  u_i=\sum_{c=0}^4\epsilon_c\widetilde u_i^{(c)}\in\ker_{R_{15}}A,
  \qquad u_i\bmod g_c=u_i^{(c)},\qquad i\in\{3,4\},
  \label{eq:s-lp15-packet-gluing}
\end{equation}
with every product now reduced modulo $x^{15}+1$. Thus the packet
normalization gives $(u_i)_j=\delta_{i,j}$ for $i,j\in\{3,4\}$ in
$R_{15}$, in particular $(u_3)_3=(u_4)_4=1$. For example,
\begin{align}
  (u_3)_0&=\epsilon_0+(1+x)\epsilon_1+(1+x^3)\epsilon_2
  +(x+x^2)\epsilon_3+(1+x^2)\epsilon_4
  =p_{2,3,4,7,9,11,14},\nonumber\\
  (u_4)_0&=\epsilon_0+x\epsilon_1+(x^2+x^3)\epsilon_2
  +(1+x+x^2)\epsilon_3+x^3\epsilon_4
  =p_{1,2,4,7,8,11,14}.
  \label{eq:s-lp15-gluing-entries}
\end{align}
The resulting global column vectors are
\begin{equation}
  u_3=
  \begin{pmatrix}
    p_{2,3,4,7,9,11,14}\\p_{0,1,2,3,10,11,12,14}\\p_{1,4,7,9,10,12}\\1\\0
  \end{pmatrix},\qquad
  u_4=
  \begin{pmatrix}
    p_{1,2,4,7,8,11,14}\\p_{3,5,7,10,11,14}\\p_{0,1,2,3,4,5,8,10}\\0\\1
  \end{pmatrix}.
  \label{eq:s-lp15-generators}
\end{equation}
For $U=(u_3\ u_4)$, direct multiplication verifies $AU=0$ and
$U_{\{3,4\},:}=I_2$. Thus
$q_{i,j}=[(u_i\otimes\varepsilon_j,0)]$, $i,j\in\{3,4\}$,
has cyclic orbit basis $\bar q_{i,j;a}^{\mathrm{cyc}}=x^a q_{i,j}$,
$a=0,\ldots,14$, paired with
$x^a(\varepsilon_i\otimes u_j,0)$ in the $X$ sector.
The CRT power basis consists of
\begin{equation}
  \bar q_{i,j;b}^{\mathrm{CRT}(c)}=\epsilon_cx^bq_{i,j},
  \qquad 0\le b<\deg g_c.
  \label{eq:s-lp15-bases}
\end{equation}
Using the ordering and coefficient convention of
Eq.~\eqref{eq:s-basis-change-rule}, the forward and inverse matrices
are explicitly
\begingroup
\setlength{\arraycolsep}{3pt}
\begin{equation}
  T_{15}=\left(
    \begin{array}{ccccccccccccccc}
      1 & 1 & 1 & 1 & 1 & 1 & 1 & 1 & 1 & 1 & 1 & 1 & 1 & 1 & 1 \\
      \hline
      0 & 1 & 1 & 0 & 1 & 1 & 0 & 1 & 1 & 0 & 1 & 1 & 0 & 1 & 1 \\
      1 & 0 & 1 & 1 & 0 & 1 & 1 & 0 & 1 & 1 & 0 & 1 & 1 & 0 & 1 \\
      \hline
      0 & 1 & 1 & 1 & 1 & 0 & 1 & 0 & 1 & 1 & 0 & 0 & 1 & 0 & 0 \\
      0 & 0 & 1 & 1 & 1 & 1 & 0 & 1 & 0 & 1 & 1 & 0 & 0 & 1 & 0 \\
      0 & 0 & 0 & 1 & 1 & 1 & 1 & 0 & 1 & 0 & 1 & 1 & 0 & 0 & 1 \\
      1 & 0 & 0 & 0 & 1 & 1 & 1 & 1 & 0 & 1 & 0 & 1 & 1 & 0 & 0 \\
      \hline
      0 & 0 & 0 & 1 & 0 & 0 & 1 & 1 & 0 & 1 & 0 & 1 & 1 & 1 & 1 \\
      1 & 0 & 0 & 0 & 1 & 0 & 0 & 1 & 1 & 0 & 1 & 0 & 1 & 1 & 1 \\
      1 & 1 & 0 & 0 & 0 & 1 & 0 & 0 & 1 & 1 & 0 & 1 & 0 & 1 & 1 \\
      1 & 1 & 1 & 0 & 0 & 0 & 1 & 0 & 0 & 1 & 1 & 0 & 1 & 0 & 1 \\
      \hline
      0 & 1 & 1 & 1 & 1 & 0 & 1 & 1 & 1 & 1 & 0 & 1 & 1 & 1 & 1 \\
      1 & 0 & 1 & 1 & 1 & 1 & 0 & 1 & 1 & 1 & 1 & 0 & 1 & 1 & 1 \\
      1 & 1 & 0 & 1 & 1 & 1 & 1 & 0 & 1 & 1 & 1 & 1 & 0 & 1 & 1 \\
      1 & 1 & 1 & 0 & 1 & 1 & 1 & 1 & 0 & 1 & 1 & 1 & 1 & 0 & 1
  \end{array}\right),\qquad
  T_{15}^{-1}=\left(
    \begin{array}{c|cc|cccc|cccc|cccc}
      1 & 1 & 0 & 1 & 0 & 0 & 0 & 1 & 0 & 0 & 0 & 1 & 0 & 0 & 0 \\
      1 & 0 & 1 & 0 & 1 & 0 & 0 & 0 & 1 & 0 & 0 & 0 & 1 & 0 & 0 \\
      1 & 1 & 1 & 0 & 0 & 1 & 0 & 0 & 0 & 1 & 0 & 0 & 0 & 1 & 0 \\
      1 & 1 & 0 & 0 & 0 & 0 & 1 & 0 & 0 & 0 & 1 & 0 & 0 & 0 & 1 \\
      1 & 0 & 1 & 1 & 1 & 0 & 0 & 1 & 0 & 0 & 1 & 1 & 1 & 1 & 1 \\
      1 & 1 & 1 & 0 & 1 & 1 & 0 & 1 & 1 & 0 & 1 & 1 & 0 & 0 & 0 \\
      1 & 1 & 0 & 0 & 0 & 1 & 1 & 1 & 1 & 1 & 1 & 0 & 1 & 0 & 0 \\
      1 & 0 & 1 & 1 & 1 & 0 & 1 & 1 & 1 & 1 & 0 & 0 & 0 & 1 & 0 \\
      1 & 1 & 1 & 1 & 0 & 1 & 0 & 0 & 1 & 1 & 1 & 0 & 0 & 0 & 1 \\
      1 & 1 & 0 & 0 & 1 & 0 & 1 & 1 & 0 & 1 & 0 & 1 & 1 & 1 & 1 \\
      1 & 0 & 1 & 1 & 1 & 1 & 0 & 0 & 1 & 0 & 1 & 1 & 0 & 0 & 0 \\
      1 & 1 & 1 & 0 & 1 & 1 & 1 & 1 & 0 & 1 & 1 & 0 & 1 & 0 & 0 \\
      1 & 1 & 0 & 1 & 1 & 1 & 1 & 1 & 1 & 0 & 0 & 0 & 0 & 1 & 0 \\
      1 & 0 & 1 & 1 & 0 & 1 & 1 & 0 & 1 & 1 & 0 & 0 & 0 & 0 & 1 \\
      1 & 1 & 1 & 1 & 0 & 0 & 1 & 0 & 0 & 1 & 1 & 1 & 1 & 1 & 1
  \end{array}\right).
  \label{eq:s-lp15-basis-change-matrices}
\end{equation}
\endgroup
The horizontal blocks have sizes $1,2,4,4,4$. The vertical blocks in
the inverse have the same sizes. Thus the first column of $T_{15}$
refers to cyclic coordinate $a=0$, while its first row refers to
$(c,b)=(0,0)$. In particular,
\begin{align}
  \bar q_{i,j;0}^{\mathrm{CRT}(2)}
  &=\sum_{a\in\{1,2,3,4,6,8,9,12\}}\bar q_{i,j;a}^{\mathrm{cyc}},\nonumber\\
  \bar q_{i,j;0}^{\mathrm{cyc}}&=\sum_{c=0}^4\bar q_{i,j;0}^{\mathrm{CRT}(c)}.
  \label{eq:s-lp15-basis-change-examples}
\end{align}
The dual $X$ rows transform by $T_{15}^{-\mathsf T}$, preserving the
identity matrix of Pauli pairings. Applying $T_{15}$ to both Pauli sectors
would not give paired bases. The reciprocal factors $g_2,g_3$ are
handled by this same rule using the inverse transpose.
The component $E_0$ contributes $4^2+2^2=20$ logical pairs, of which
four are canonical. Each of the other components contributes
$4\deg g_c$. Consequently the complete basis has
$60+16=76$ pairs, with 12 residual pairs in the left tensor sector
and four in the right tensor sector.

\subsubsection{Algebraic surgery construction and thickening}
For the full fibre $(3,3)$, set
\begin{equation}
  H_L=\varepsilon_4^{\mathsf T},\quad
  A'=
  \begin{pmatrix}A\\H_L
  \end{pmatrix},\quad
  \ker A'=R_{15}u_3\oplus K_{\mathrm{res}},
  \label{eq:s-lp15-carving}
\end{equation}
where
$K_{\mathrm{res}}=\operatorname{span}_{\mathbb F_2}
\{\epsilon_0(\varepsilon_0+\varepsilon_1),
\epsilon_0(\varepsilon_0+\varepsilon_2)\}$.
This kernel has binary dimension 17. Attaching it to the cokernel
coordinate $\varepsilon_3$ measures the 15 canonical operators together
with the two residual operators represented by
$(\epsilon_0(\varepsilon_0+\varepsilon_r)\otimes\varepsilon_3,0)$,
$r=1,2$.

Packet selection is explicit at the same address. For
$S\subseteq\{0,1,2,3,4\}$, define
\begin{equation}
  H_{L,S}=
  \begin{pmatrix}
    \varepsilon_4^{\mathsf T}\\
    g_S\varepsilon_0^{\mathsf T}\\
    g_S\varepsilon_1^{\mathsf T}\\
    g_S\varepsilon_2^{\mathsf T}\\
    g_S\varepsilon_3^{\mathsf T}
  \end{pmatrix},\qquad
  g_S=\prod_{c\in S}g_c,\qquad e_S=\sum_{c\in S}\epsilon_c.
  \label{eq:s-lp15-packet-carving}
\end{equation}
Then $\ker(A;H_{L,S})=R_{15}e_Su_3\oplus e_SK_{\mathrm{res}}$.
For example, $S=\{2,3\}$ gives
$g_S=1+x+x^3+x^4+x^5+x^7+x^8$ and exactly eight canonical
readouts, with no residual readout. The five singleton selections give
one, two, four, four, and four canonical readouts, respectively. Only
$S=\{0\}$ also retains the two residuals. If exact canonical selection
is required when $0\in S$, the additional rows
$\epsilon_0\varepsilon_1^{\mathsf T}$ and
$\epsilon_0\varepsilon_2^{\mathsf T}$ remove these residuals while
annihilating $u_3$. These statements specify the measured subspace.
Distance claims below refer to the factor for the full fibre in
Eq.~\eqref{eq:s-lp15-carving}.

The chosen CRT outcomes can be obtained explicitly in the thickened
construction. For a normalized tube generator $t$ with $t_q=1$, set $\xi_b^{(c)}=(\epsilon_cx^bu_i)\otimes t$ with $c\in S$ and $0\le b<d_c$.
Then $\partial_1\xi_b^{(c)}=0$ and
$f_1\xi_b^{(c)}=(\epsilon_cx^bu_i\otimes\varepsilon_j,0)$.
Multiplying the added $Z$ checks with binary coefficients given by the
lifted column of $\xi_b^{(c)}$ cancels all auxiliary support and yields
exactly this representative of $\bar q_{i,j;b}^{\mathrm{CRT}(c)}$.
Thus each displayed CRT operator is an available outcome parity.
It need not be an individual added check or an element of the raw
kernel basis. This also explains why the same $H_L$ remains compatible
with every invertible change of basis within its measured space.

For the thickening, we choose a tube with three monomials per row,
\begin{equation}
  D=
  \begin{pmatrix}
    1&1&1&0\\
    0&x^9&x^{10}&1\\
    x^{11}&0&x^4&1
  \end{pmatrix},\qquad
  t=
  \begin{pmatrix}1\\h\\1+h\\x^{10}+(x^9+x^{10})h
  \end{pmatrix},\qquad
  h=p_{0,3,5,7,8,9,10,11,12}.
  \label{eq:s-lp15-tube}
\end{equation}
The minor obtained by deleting column zero is
$\Delta=x^4+x^9+x^{10}$, a unit in $R_{15}$.
The first two equations give $z_2=z_0+z_1$ and
$z_3=x^{10}z_0+(x^9+x^{10})z_1$. The third gives
$\Delta z_1=(x^4+x^{10}+x^{11})z_0$.
Thus $h=\Delta^{-1}(x^4+x^{10}+x^{11})$ and
$\ker D=R_{15}t$, normalized at $q=0$ by $t_0=1$.
Enumeration of all $2^{15}-1=32767$ nonzero kernel words gives a
classical $[60,15,11]$ code, with 15 words of weight eleven.
A word of minimum weight is
$\Delta t=(\Delta,x^4+x^{10}+x^{11},x^9+x^{11},x^5+x^6+x^{13})^{\mathsf T}$.
As at lift nine, target orders $123$, $234$, and $341$, counting from one,
give three gate rounds without conflicts.

\subsubsection{Adapter between two blocks}
Take two copies of Eq.~\eqref{eq:s-lp15-seed}. Carve fibre $(3,3)$
in the first and $(4,4)$ in the second, and append a link of offset
$\alpha=1$:
\begin{equation}
  A'_{\mathrm{ad}}=
  \begin{pmatrix}
    A&0\\ \varepsilon_4^{\mathsf T}&0\\
    0&A\\0&\varepsilon_3^{\mathsf T}\\
    \varepsilon_3^{\mathsf T}&x\varepsilon_4^{\mathsf T}
  \end{pmatrix},\qquad
  \ker A'_{\mathrm{ad}}=
  R_{15}
  \begin{pmatrix}u_3^{[1]}\\x^{-1}u_4^{[2]}
  \end{pmatrix}
  \oplus K_{\mathrm{res},1}\oplus K_{\mathrm{res},2}.
  \label{eq:s-lp15-adapter}
\end{equation}
This is a $9\times10$ factor. Its auxiliary boundaries are
\begin{equation}
  \partial_1=
  \begin{pmatrix}A'_{\mathrm{ad}}\otimes I_4\\I_{10}\otimes D
  \end{pmatrix},
  \qquad
  \partial_0=(I_9\otimes D\mid A'_{\mathrm{ad}}\otimes I_3),
  \label{eq:s-lp15-adapter-boundaries}
\end{equation}
of sizes $66\times40$ and $27\times66$.
For explicit attachment maps, put
$F_1^{[\mu]}=\binom{I_5\otimes\varepsilon_{j_\mu}}{0_{9\times5}}$
and $F_0^{[\mu]}=(I_3\mid0_{3\times1})\otimes\varepsilon_{j_\mu}$,
with $(j_1,j_2)=(3,4)$. Then
\begin{align}
  f_1&=(F_1^{[1]}\oplus F_1^{[2]})\otimes e_0^{\mathsf T},\nonumber\\
  f_0&=\bigl((F_0^{[1]}\oplus F_0^{[2]}\mid0_{30\times1})
  \otimes e_0^{\mathsf T}\mid0_{30\times30}\bigr).
  \label{eq:s-lp15-adapter-wiring}
\end{align}
The maps have sizes $68\times40$ and $30\times66$. Their chain
identity verifies commutation after insertion into the merged checks.
The 15 canonical products are
\begin{equation}
  P_a^{\mathrm{cyc}}=\bar Z_{33;a}^{[1],\mathrm{cyc}}\bar Z_{44;a-1}^{[2],\mathrm{cyc}},
  \qquad a\in\mathbb Z_{15}.
  \label{eq:s-lp15-adapter-products}
\end{equation}
Indeed, the combination of added checks specified by
$x^a(u_3^{[1]},x^{-1}u_4^{[2]})\otimes t$ cancels all auxiliary support and
returns $P_a^{\mathrm{cyc}}$. Four independent local residual operators are also
measured. The complete measured subspace has dimension 19 and
intersects either individual canonical register trivially. Its
intersection with either complete data register consists of that
block's two residual readouts.
The adapter uses $15(9\cdot4+10\cdot3)=990$ auxiliary qubits and
2010 qubits in total. The merged check ranks are 843 and 1034, giving
$k=133=2\cdot76-19$.

\subsection{$[[1122,148,\leq 20]]$ LP data code with $l=33$}
\label{app:lp33-timing}
The largest timing instance uses $R_{33}$, $B=A^\dagger$, and
\begin{equation}
  A=
  \begin{pmatrix}
    1&1&1&1&1\\
    1&x^{14}&x^{19}&x^{11}&x^{26}\\
    1&x^{13}&x^2&x^{15}&x^{21}
  \end{pmatrix},\qquad
  D=
  \begin{pmatrix}
    x^{27}&x^{19}&x^{13}&1&0\\
    0&x^{22}&x^8&x^{13}&x^{20}\\
    x^8&0&x^{29}&x^{22}&x^{20}\\
    x^{19}&x^{29}&0&x^{12}&x^{20}
  \end{pmatrix}.
  \label{eq:s-lp33-seed-tube}
\end{equation}
The target fibre is $(i,j)=(3,3)$ with coordinates numbered from zero, $H_L=\varepsilon_4^{\mathsf T}$, and $A'=\binom{A}{H_L}$. The coupled tube coordinate is $q=2$. The binary lift of the $4\times4$ minor of $D$ obtained by deleting column two has rank 132, so that minor is invertible over $R_{33}$. Consequently, for any choice of $z_2\in R_{33}$, the other four entries of $z\in\ker D$ are uniquely determined. Taking $z_2=1$ defines $t$ with $\ker D=R_{33}t$ and $t_2=1$, as required by the coordinate attachment.

The auxiliary has $33(4\cdot5+5\cdot4)=1320$ qubits. Together with the $33(25+9)=1122$ data qubits this gives 2442. Its merged $X$ and $Z$ matrices have shapes $1023\times2442$ and $1320\times2442$ and binary ranks 1017 and 1312. Direct multiplication verifies their commutation and gives $k=2442-1017-1312=113$. The same matrix check gives ranks $(237,307)$ for lift nine and $(399,517)$ for lift fifteen, agreeing with the preceding constructions. The fibre protocol at lift 33 has 33 canonical and two residual readouts, reducing $148$ data logical qubits to $113$. These rank checks establish commutation and dimension. They are separate from the distance upper bound used to identify instance III and from the circuit-level benchmark below.

\subsubsection{Logical error rates}
\label{app:lp33-ler}
\paragraph{Circuit simulations.}
We compare idling with $X$-type fibre surgery on the $[[1122,148,d\leq20]]$ data code using 20 syndrome extraction rounds.
The physical error probabilities are $p=0.0016$, $0.002$, $0.0025$, and
$0.003$. The shared circuit construction and decoder parameters are given
in \hyperref[app:simulation-methods]{Simulation methods and decoder settings}.
Memory tracks 148 logical observables in either basis. Surgery records $148+33$ observables in the $X$ experiment and $148-(33+2)$ surviving
logical observables in the $Z$ experiment. The two additional measured
residual observables are not part of the canonical basis.

The circuit-level benchmark compares idling with the full surgery protocol using the average of the logical error probabilities in the $X$ and $Z$ bases with equal weights,
reported as an effective LER per cycle. Figure~\ref{fig:ler} shows the
results for each basis.
The fitted effective distances are $d_{\mathrm{eff}}=15.9^{+2.9}_{-2.8}$
for idling and $22.8^{+1.7}_{-1.6}$ for surgery (95\% confidence bounds).
The surgery average rises more steeply over the sampled range but also departs
from a single power law. These fit parameters characterize the present
construction and decoder over this range. They do not establish the code
distance or a threshold.

The analysis defines $\bar P=(P_X+P_Z)/2$ for experiments with 20 rounds and reports $r=[1-(1-2\bar P)^{1/20}]/2$. The slope fits use $p=0.002,0.0025,0.003$ and $\bar P=C(p/0.0025)^b$, with $d_{\rm eff}=2b-1$. The approximate $p$-value for the fit to the surgery data is $0.008$. Shot counts are given below. These circuit results do not validate the shuttling model or its excluded motion mechanisms.

Table~\ref{tab:circuit-errors} gives the counts for each basis underlying
Fig.~\ref{fig:ler} in the main text. These are counts per full shot.
The table does not normalize by the number of cycles or logical qubits.

\begin{table}[htbp]
  \caption{\textbf{Circuit-level sample sizes and logical failures.}
    The stage columns partition the total errors by where the output was
    accepted. The last three columns independently partition the same total
    into mutually exclusive error types. ``Both'' means a block logical and a
    time-like readout are wrong in the same shot. At $p=0.0016$, all reported
  errors arise in T2 BP--LSD. T1 contributes no logical failures.}
  \label{tab:circuit-errors}
  \centering
  \small
  \setlength{\tabcolsep}{5pt}
  \begin{tabular}{llrrrrrrr}
    \hline\hline
    $p$ & Experiment & Shots & \multicolumn{3}{c}{Errors by decoder stage} & \multicolumn{3}{c}{Errors by type} \\
    \cline{4-6}\cline{7-9}
    & & & T1 & T2 & Total & Block only & Time-like only & Both \\
    \hline
    0.0016 & Idling $X$ & 5\,000\,000 & 0 & 2 & 2 & 2 & 0 & 0 \\
    0.0016 & Idling $Z$ & 5\,000\,000 & 0 & 1 & 1 & 1 & 0 & 0 \\
    0.0016 & Surgery $X$ & 5\,000\,000 & 0 & 1 & 1 & 0 & 1 & 0 \\
    0.0016 & Surgery $Z$ & 5\,000\,000 & 0 & 4 & 4 & 4 & 0 & 0 \\
    \hline
    0.002 & Idling $X$ & 5\,000\,000 & 0 & 6 & 6 & 6 & 0 & 0 \\
    0.002 & Idling $Z$ & 5\,000\,000 & 0 & 10 & 10 & 10 & 0 & 0 \\
    0.002 & Surgery $X$ & 4\,000\,000 & 3 & 2 & 5 & 0 & 3 & 2 \\
    0.002 & Surgery $Z$ & 7\,000\,000 & 0 & 29 & 29 & 29 & 0 & 0 \\
    \hline
    0.0025 & Idling $X$ & 3\,000\,000 & 0 & 29 & 29 & 29 & 0 & 0 \\
    0.0025 & Idling $Z$ & 3\,000\,000 & 0 & 23 & 23 & 23 & 0 & 0 \\
    0.0025 & Surgery $X$ & 2\,775\,000 & 16 & 8 & 24 & 5 & 16 & 3 \\
    0.0025 & Surgery $Z$ & 3\,000\,000 & 0 & 115 & 115 & 115 & 0 & 0 \\
    \hline
    0.003 & Idling $X$ & 450\,000 & 1 & 16 & 17 & 17 & 0 & 0 \\
    0.003 & Idling $Z$ & 500\,000 & 0 & 27 & 27 & 27 & 0 & 0 \\
    0.003 & Surgery $X$ & 500\,000 & 8 & 9 & 17 & 2 & 8 & 7 \\
    0.003 & Surgery $Z$ & 500\,000 & 0 & 231 & 231 & 231 & 0 & 0 \\
    \hline\hline
  \end{tabular}
\end{table}

For idling in either basis and surgery in the $Z$ basis, all recorded
logical observables are block observables, so every error is a failure of block observables only. Surgery in the $X$ basis has 148 block observables followed by
33 time-like readouts. We classify the saved masks of observable errors
using this ordering. At $p=0.002$, $0.0025$, and $0.003$, respectively,
all $3$, $16$, and $8$ T1 surgery $X$ errors are time-like only.
The corresponding T2 counts in the order (block only, time-like only,
both) are $(0,0,2)$, $(5,0,3)$, and $(2,0,7)$. The single reported
surgery $X$ error at $p=0.0016$ is time-like only and occurs in T2 BP--LSD.

\paragraph{Code-capacity benchmarks.}
\label{app:code-capacity}
Table~\ref{tab:code-capacity-counts} summarizes code-capacity results
over $p=0.015$--$0.05$ using the shared BP--LSD settings, comparing the
bare $[[1122,148]]$ LP$_{33}$ block with
the merged $[[2442,113]]$ surgery code. The reported data comprise
348 shards, 90\,336\,000 shots, and 1\,273 logical failures.
At $p=0.015$, two independent batches of $10^7$ shots per code and basis
are combined, giving $2\times10^7$ shots per entry. The first batch has
$0$, $3$, $0$, and $1$ failures for bare $X$, bare $Z$, merged $X$, and
merged $Z$, respectively. The second adds $3$, $0$, $0$, and $0$.
The combined LER estimates are therefore $1.5\times10^{-7}$ in both
bases of the bare code and $5.0\times10^{-8}$ for merged $Z$. Merged $X$ has
zero failures, giving a one-sided 95\% upper bound of $1.5\times10^{-7}$.
These small failure counts leave substantial statistical uncertainty.

At $p=0.03$--$0.05$, the merged code has a lower measured LER in the $X$
basis, whereas its $Z$-basis LER exceeds the bare code's at $p=0.04$
and $0.05$. These results for finite codes do not establish a threshold.

\begin{table}[htbp]
  \caption{\textbf{Code-capacity shot counts and logical error rates for $p=0.015$--$0.05$.}
    Each error is a block failure over a full shot in the measured basis. All
    shots use direct BP--LSD decoding, with no Relay-BP stage or time-like
    readout. The $p=0.015$ entries combine both independent batches.
    Nonzero LER entries are observed errors divided by shots. Entries
    preceded by $<$ are one-sided 95\% upper confidence bounds after zero
  failures, $1-0.05^{1/N}$. No normalization by the number of cycles is used.}
  \label{tab:code-capacity-counts}
  \centering
  \small
  \setlength{\tabcolsep}{5pt}
  \begin{tabular}{llrrrrrr}
    \hline\hline
    $p$ & Code & \multicolumn{3}{c}{$X$ basis} & \multicolumn{3}{c}{$Z$ basis} \\
    \cline{3-5}\cline{6-8}
    & & Shots & Errors & LER & Shots & Errors & LER \\
    \hline
    0.015 & Bare & 20\,000\,000 & 3 & $1.5\times10^{-7}$ & 20\,000\,000 & 3 & $1.5\times10^{-7}$ \\
    0.015 & Merged & 20\,000\,000 & 0 & $<1.5\times10^{-7}$ & 20\,000\,000 & 1 & $5\times10^{-8}$ \\
    \hline
    0.02 & Bare & 2\,000\,000 & 1 & $5\times10^{-7}$ & 2\,000\,000 & 2 & $1\times10^{-6}$ \\
    0.02 & Merged & 2\,000\,000 & 1 & $5\times10^{-7}$ & 2\,000\,000 & 3 & $1.5\times10^{-6}$ \\
    \hline
    0.03 & Bare & 500\,000 & 26 & $5.2\times10^{-5}$ & 500\,000 & 30 & $6\times10^{-5}$ \\
    0.03 & Merged & 500\,000 & 2 & $4\times10^{-6}$ & 500\,000 & 38 & $7.6\times10^{-5}$ \\
    \hline
    0.04 & Bare & 84\,000 & 101 & $1.2\times10^{-3}$ & 84\,000 & 102 & $1.21\times10^{-3}$ \\
    0.04 & Merged & 100\,000 & 4 & $4\times10^{-5}$ & 28\,000 & 105 & $3.75\times10^{-3}$ \\
    \hline
    0.05 & Bare & 8\,000 & 193 & $2.41\times10^{-2}$ & 8\,000 & 210 & $2.62\times10^{-2}$ \\
    0.05 & Merged & 20\,000 & 11 & $5.5\times10^{-4}$ & 4\,000 & 437 & $1.09\times10^{-1}$ \\
    \hline\hline
  \end{tabular}
\end{table}

\section{Distance estimates for merged codes from QDistRnd}
\label{app:merged-distance-confidence}

QDistRnd~\cite{PryadkoShabashovKozin2022} finds logical operators of minimum
sampled weights $(9,8)$, $(11,10)$, and $(21,20)$ in the $(X,Z)$ sectors
of the three merged codes for full fibres. These searches support distance
estimates of $8$, $10$, and $20$, respectively
(Table~\ref{tab:merged-distance-empirical-scores}). Each found logical
operator gives an upper bound on the true sector distance. QDistRnd alone
cannot exclude an undiscovered logical operator of smaller weight.

\begin{table}[htbp]
  \caption{\textbf{Distance results for merged codes from QDistRnd.}
    The smallest sampled logical weights $\widehat d_X$ and $\widehat d_Z$
    give the distance estimate for the merged code
    $\widehat d=\min(\widehat d_X,\widehat d_Z)$ and the guaranteed upper
    bound $d\le\widehat d$. The last column is the combined empirical
    score for both sector estimates. The counts $N_P,m_P,h_P$ and mean
    multiplicity $\mu_P=h_P/m_P$ are defined below. Multiplicities are
    rounded. The $l=33$ $Z$ row is the last saved checkpoint before
  the seven-day timeout.}
  \label{tab:merged-distance-empirical-scores}
  \centering
  \setlength{\tabcolsep}{5pt}
  \begin{tabular}{rrcrrrrrrr}
    \hline\hline
    Lift $l$ & Merged $[[n,k]]$ & $P$ & $\widehat d_P$ & $N_P$ & $m_P$ & $h_P$
    & $\mu_P$ & $\widehat d$ & $C_{XZ}$ (\%) \\
    \hline
    9  & $[[585,41]]$   & $X$ & 9  & 1\,000       & 9   & 82  & 9.1111 & 8  & 99.95 \\
    &                & $Z$ & 8  & 1\,000       & 100 & 784 & 7.8400 &    &       \\
    \hline
    15 & $[[975,59]]$   & $X$ & 11 & 1\,566       & 29  & 154 & 5.3103 & 10 & 99.01 \\
    &                & $Z$ & 10 & 2\,967       & 100 & 531 & 5.3100 &    &       \\
    \hline
    33 & $[[2442,113]]$ & $X$ & 21 & 986\,948     & 33  & 175 & 5.3030 & 20 & 97.60 \\
    &                & $Z$ & 20 & 6\,823\,800 & 100 & 396 & 3.9600 &    &       \\
    \hline\hline
  \end{tabular}
\end{table}

For each sector $P\in\{X,Z\}$, $N_P$ is the number of random information
sets (trials), $m_P$ is the number of distinct tracked logical words at
the smallest sampled weight, and $h_P$ is the total number of times those
tracked words were found, including repeated discoveries. Their mean
multiplicity is $\mu_P=h_P/m_P$, the average number of discoveries per tracked word.
For example, at $l=33$ in the $Z$ sector, the last saved checkpoint
records $6\,823\,800$ trials and $396$ discoveries among $100$ tracked
words, so $\mu_Z=3.96$.
We use the package's empirical score $C_P=1-e^{-\mu_P}$ and combine the
sectors as $C_{XZ}=\max\{0,1-e^{-\mu_X}-e^{-\mu_Z}\}$.
Under the model's assumptions that words of equal weight are equally
discoverable and words of smaller weight are at least as discoverable, a larger score gives stronger evidence that no smaller
logical operator was missed. These are heuristic scores, not calibrated
confidence levels or proofs of equality. Thus the estimates $d=8$ and
$d=10$ exceed the $99\%$ empirical target. The estimate $d=20$ has
$97.60\%$ combined empirical support at this checkpoint. The $l=33$
$Z$ search reached its seven-day timeout on September 27, 2026,
before reaching the $99\%$ combined target.

The runs use QDistRnd 0.9.5 under GAP 4.15.1, $\mathbb F_2$, seed
$260920$, and \texttt{mindist=0}. The extended $l=15,33$ searches use
the stopping threshold \texttt{maxav=5.3} for each sector's mean
multiplicity. The $l=15$ sectors and the $l=33$ $X$ sector reached this
threshold. The $l=33$ $Z$ search stopped at the timeout.
QDistRnd tracks at most $100$ distinct words of minimum weight, and only
discoveries of those tracked words enter the score. Earlier prefixes from the same seed are counted once.

\section{Simulation methods and decoder settings}
\label{app:simulation-methods}
The following setup and decoder parameters are shared by the circuit
experiments. Each experiment's number of rounds, physical error
probabilities, recorded logical observables, and results are given with
the corresponding code example.

\subsection{Circuit construction}
We implement the high-rate protocol for measuring Pauli products with a compiler that translates the merged CSS check matrices and batch readout vectors into a Stim~\cite{gidney2021stim} circuit. The simulations use $X$-type surgery, although the formulas in the main text are written for $Z$-type surgery.
Data qubits are initialized in the experiment basis ($X$ or $Z$), and auxiliary qubits are initialized in the $Z$ basis. After the chosen
number of syndrome extraction rounds, the auxiliary qubits are measured
in $Z$ to detach them, followed by data readout in the experiment basis.
Stim \texttt{DETECTOR} annotations compare successive check outcomes,
with boundary detectors for checks fixed by preparation or final readout.
In the $X$ experiment, \texttt{OBSERVABLE\_INCLUDE} annotations record
logicals from data measurements and surgery observables obtained from
products of selected stabilizer outcomes from the first round.
In the $Z$ experiment, \texttt{OBSERVABLE\_INCLUDE} includes only logical
observables that survive the corresponding surgery.
Detectors are added only in the stabilizer basis corresponding to the
experiment. Gate depolarization and preparation and readout flips use
probability $p$. Stim's compiled detector sampler supplies detector events
and observable flips for decoding.

Both surgery and memory use the \texttt{EdgeColoring} schedule for syndrome extraction in the QLDPC package~\cite{perlin2023qldpc}. We greedily color the edges of each CSS Tanner subgraph, process the $X$-check subgraph before the
$Z$-check subgraph, and apply the controlled-Pauli gates one color at a time. Gates of the same color share no qubits and form a parallel layer. This coloring
schedule is used in all reported circuit-level LER simulations.

\subsection{Decoding}
The circuit-level simulations use two decoding methods in sequence:
T1 is Relay-BP~\cite{roffe_decoding_2020,mullerImprovedBeliefPropagation2025b,hillmannLocalizedStatisticsDecoding2025}, including its initial plain BP iterations, and T2 is the
BP-LSD fallback applied only to shots for which Relay-BP does not converge.
Table~\ref{tab:ler-decoder-parameters} lists the shared decoder parameters and settings for
idling and surgery in both measurement bases. The number of rounds and
physical error probabilities are specified with the corresponding code example. Both methods decode the
same undecomposed detector error model (DEM), using the
probabilities of its error mechanisms as priors. T2 starts a fresh BP-LSD decode
of the original detector syndrome. It does not inherit Relay-BP messages. The decoder parameters were tuned for convergence on these runs.

\begin{table}[htbp]
  \caption{\textbf{Shared parameters for the circuit simulations of logical error rates.}
    T1 denotes the complete Relay-BP call, and T2 denotes BP--LSD on its
    nonconverged shots. The initial plain BP iterations are part of T1.
    Relay-BP stops at the first converged solution. Its seed controls decoder
    randomness, not sampling of circuit noise. The BP--LSD settings also apply
    to the direct code-capacity runs, which omit T1 and use
  error probabilities of physical qubits as priors.}
  \label{tab:ler-decoder-parameters}
  \centering
  \small
  \setlength{\tabcolsep}{4pt}
  \renewcommand{\arraystretch}{1.1}
  \begin{tabular}{p{0.30\linewidth}p{0.40\linewidth}p{0.23\linewidth}}
    \hline\hline
    Parameter & Implementation keyword & Setting \\
    \hline
    \multicolumn{3}{l}{\textit{Shared circuit settings}} \\
    Measurement bases & \texttt{basis} & $X$ and $Z$, separately \\
    DEM hyperedge decomposition & \texttt{decompose\_errors} & \texttt{False} \\
    \hline
    \multicolumn{3}{l}{\textit{T1: Relay-BP}} \\
    Decoder implementation & \texttt{name} & \texttt{RelayDecoderF32} \\
    Initial plain BP iteration limit & \texttt{pre\_iter} & 120 \\
    Maximum relay sets & \texttt{num\_sets} & 30 \\
    Iteration limit per relay set & \texttt{set\_max\_iter} & 60 \\
    Min-sum scaling $\alpha$ & \texttt{alpha} & $0.85$ \\
    Iteration scaling factor & \texttt{alpha\_iteration\_scaling\_factor} & $1.0$ \\
    Initial memory strength $\gamma_0$ & \texttt{gamma0} & $0.1$ \\
    Sampling interval for memory strength & \texttt{gamma\_dist\_interval} & $(0.1,0.8)$ \\
    Converged solutions before stopping & \texttt{stop\_nconv} & 1 \\
    Decoder random seed & \texttt{seed} & 0 \\
    \hline
    \multicolumn{3}{l}{\textit{T2: BP--LSD fallback}} \\
    BP update rule & \texttt{bp\_method} & \texttt{minimum\_sum} \\
    BP update schedule & \texttt{schedule} & \texttt{serial} \\
    Maximum BP iterations & \texttt{max\_iter} & 100 \\
    Min-sum scaling factor & \texttt{ms\_scaling\_factor} & $0.625$ \\
    LSD method & \texttt{lsd\_method} & \texttt{lsd\_e} (LSD-E) \\
    LSD order & \texttt{lsd\_order} & 5 \\
    OpenMP threads per decoder & \texttt{omp\_thread\_count} & 1 \\
    \hline\hline
  \end{tabular}
\end{table}

T1 errors are converged but
logically incorrect Relay-BP outputs. T2 errors are logical failures
after BP--LSD decoding of shots not resolved by Relay-BP. Each failed
shot contributes to exactly one of these stage totals, so
$E_{\rm total}=E_{\rm T1}+E_{\rm T2}$.

\subsection{Code-capacity simulations}
Code-capacity runs use direct BP--LSD decoding with the settings in
Table~\ref{tab:ler-decoder-parameters}, including min-sum scaling $0.625$.
They omit Relay-BP and use physical qubit error probabilities as priors.
A failure is an error in any block logical observable in the measured
basis. These runs have no time-like surgery readouts.

\end{document}